\documentclass[10pt, conference, letterpaper]{IEEEtran}
\usepackage{cite}
\usepackage[caption=false,font=footnotesize]{subfig}
\usepackage{graphicx}
\usepackage{multirow}
\usepackage{textcomp}
\usepackage{gensymb}
\usepackage{booktabs}
\usepackage{amsmath,amssymb,amsfonts}

\usepackage[ruled,linesnumbered]{algorithm2e}
\usepackage[noend]{algpseudocode}
\def\BibTeX{{\rm B\kern-.05em{\sc i\kern-.025em b}\kern-.08em
    T\kern-.1667em\lower.7ex\hbox{E}\kern-.125emX}}
\begin{document}

\title{Deep Learning-based Modulation Classification of Practical OFDM Signals for Spectrum Sensing}

\author{\IEEEauthorblockN{Byungjun Kim\IEEEauthorrefmark{1}, Christoph Mecklenbr{\"a}uker\IEEEauthorrefmark{3}, Peter Gerstoft\IEEEauthorrefmark{1}\IEEEauthorrefmark{2}}
\IEEEauthorblockA{\IEEEauthorrefmark{1}\textit{Dept. of ECE, University of California, San Diego, La Jolla, USA} \\
\IEEEauthorblockA{\IEEEauthorrefmark{2}\textit{Marine Physical Laboratory, University of California, San Diego, La Jolla, USA}}
\IEEEauthorblockA{\IEEEauthorrefmark{3}\textit{Institute of Telecommunications, TU Wien, Vienna, Austria}}
}}
\maketitle
\newcommand{\Exp}[1]{\mathrm{e}^{#1}}
\newcommand{\fTX}{f_{\rm{TX}}}
\newcommand{\fRX}{f_{\rm{RX}}}
\newcommand{\TIFFT}{T_{\rm{IFFT}}}
\newcommand{\NFFT}{N_{\rm{FFT}}}
\newcommand{\TCP}{T_{\rm{CP}}}
\newcommand{\NCP}{N_{\rm{CP}}}
\newcommand{\dFSCS}{\Delta f_{\rm{SCS}}}
\newcommand{\fSDR}{f_{\rm{SDR}}}
\newcommand{\fG}{f_{\rm{5G}}}

\begin{abstract}
In this study, the modulation of symbols on OFDM subcarriers is classified for transmissions following Wi-Fi~6 and 5G downlink specifications. First, our approach estimates the OFDM symbol duration and cyclic prefix length based on the cyclic autocorrelation function. We propose a feature extraction algorithm characterizing the modulation of OFDM signals, which includes removing the effects of a synchronization error. The obtained feature is converted into a 2D histogram of phase and amplitude and this histogram is taken as input to a convolutional neural network (CNN)-based classifier. The classifier does not require prior knowledge of protocol-specific information such as Wi-Fi preamble or resource allocation of 5G physical channels. The classifier's performance, evaluated using synthetic and real-world measured over-the-air (OTA) datasets, achieves a minimum accuracy of 97\% accuracy with OTA data when SNR is above the value required for data transmission.
\end{abstract}


%
\IEEEpeerreviewmaketitle

\vspace{-5pt}
\section{Introduction}
The growth of wireless technologies in the scarce radio spectrum has strongly prioritized spectral efficiency: A challenge that is being addressed by, e.g., (massive) MIMO technology, joint radar communications, and cognitive radio \cite{Bjoernson2016,Mishra2019,Devroye2006}. Here, we focus on an essential component of cognitive radio, namely intelligent spectrum sensing, which allows for real-time characterization of radio spectrum usage and aids in online decision-making for spectrum allocation. Spectrum sensing encompasses signal detection~\cite{gao2019deep}, predicting available spectrum~\cite{yu2017spectrum}, and identifying modulation schemes. In this study, we focus on the classification of modulations of state-of-the-art wireless orthogonal frequency division multiplexing (OFDM) signals.

OFDM transmission has become foundational in current wireless communication systems, such as Wi-Fi~6 and 5G. In these systems, message bits are first encoded and subsequently mapped to digital symbols using quadrature amplitude modulation (QAM) on individual subcarriers. Many QAM symbols are modulated onto many subcarriers, so each time sample contains only a small fraction of the information carried by an OFDM symbol. As a result, the modulation classifiers designed for single-carrier signals~\cite{sathyanarayanan2023rml22, liu2023towards} are not directly applicable to OFDM signals. Therefore, an accurate modulation classifier for Wi-Fi~6 and 5G signals requires additional processing beyond using raw time-domain samples as inputs.

In contrast to a dedicated receiver (RX) as a node in a wireless network, a spectrum sensor must be able to handle OFDM signals with diverse subcarrier configurations without access to prior information about the transmission format. In Wi-Fi~6 and 5G systems, information about the user data transmission, including the modulation, is provided to the RX through a protocol-specific procedure. However, since a spectrum sensor does not have prior knowledge of the type of signals it detects, it cannot deploy the procedure to obtain user data transmission information. The parameters shaping OFDM signals, fast Fourier transform (FFT) size to generate inverse fast Fourier transform (IFFT) sequence, and cyclic prefix (CP) length, might be different even among OFDM signals with the same modulation scheme. The diverse parameter options complicate the Wi-Fi preamble structure and in the recent Wi-Fi~6 these become more diverse. This makes spectrum sensing harder only with Wi-Fi preamble to identify the modulation scheme, even though the preamble structure is known. Moreover, the carrier frequency configurations in 5G become increasingly diverse and data transmission might occupy only a part of channel bandwidth. As a result, estimation of these carrier frequency configurations is becoming increasingly difficult using transmission bandwidth and center frequency alone. Thus, a modulation classifier for spectrum sensing should estimate the modulation scheme using only the observed user data transmission without knowledge of the OFDM signal parameters including FFT size, CP length, and carrier frequency.

We propose and analyze a modulation classifier for Wi-Fi~6~\cite{ieee802.11-2021} and 5G~\cite{3GPP_38331} for a spectrum sensing system. Without knowledge of the transmitter (TX) carrier frequency, Wi-Fi preamble, or 5G control information, the classifier exploits only the basic OFDM structure, IFFT sequence, and CP. This includes the estimation of OFDM parameters: CP length and subcarrier spacing (SCS), which is directly related to the FFT size of the IFFT sequence. We focus on identifying modulation schemes used in the payload of Wi-Fi~6 signals and the physical downlink shared channel (PDSCH) of 5G signals. Signals studied in this paper are single-input single-output (SISO). For 5G, they are in the frequency range 1 (FR1), whose frequency band is below 7.125~GHz.

For the SCS and CP length estimation, the cyclic autocorrelation function (CAF) is deployed. The capability of CAF to detect intervals of repeated sequences and repetition periods enables the estimation of those parameters. We observe that symbol-level synchronization is not perfect if autocorrelation using CP is utilized only. Our preprocessing removes the effect of the synchronization error by using phase differences between phases of two adjacent OFDM symbols. The modulation classifier for Wi-Fi~6 and 5G signals should recognize high-order modulations such as 256QAM and 1024QAM since these state-of-the-art protocols include those schemes. We change the feature format to a histogram representing the distribution of the features so that the classifier can effectively capture high-order modulation characteristics.

\textbf{Related work on modulation classification:} Many papers address modulation classification for wireless communication signals~\cite{hong2019deep, al2021amc2, zhang2020automatic, gupta2020blind, pathy2021design, hong2020convolutional, liu2023towards, sathyanarayanan2023rml22, park2021deep, kumar2023automatic}. The works in~\cite{hong2019deep, al2021amc2, zhang2020automatic, gupta2020blind, pathy2021design, hong2020convolutional} study modulation classification of OFDM signals and achieve at least 78\% accuracy at 20\,dB~SNR for an AWGN channel. It is assumed that the inputs start from the first sample of the OFDM symbol duration~\cite{hong2019deep, al2021amc2, hong2020convolutional, zhang2020automatic}, which requires detecting the timing of the Wi-Fi preamble or 5G synchronization signals. To apply this approach to a spectrum sensor, the sensor needs to follow protocol-specific procedures. Further, neither of these works is evaluated on real-world measured data.

Previous works on OFDM modulation classification without symbol-level synchronization~\cite{gupta2020blind, pathy2021design, park2021deep, kumar2023automatic} and the algorithms~\cite{gupta2020blind, pathy2021design, kumar2023automatic} are evaluated with hardware-generated data. However, their algorithms~\cite{gupta2020blind,pathy2021design,kumar2023automatic} are not evaluated with high-order modulations such as 256QAM or 1024QAM, as used in Wi-Fi~6 and 5G. Moreover, since their classifier structures~\cite{gupta2020blind, pathy2021design} are designed to recognize only a fixed set of modulations, the overall structure needs to be redesigned to identify a new modulation scheme. The work~\cite{park2021deep} proposes the system to estimate SCS of OFDM signals and modulation of single-carrier signals jointly. Nonetheless, it does not estimate the modulation of OFDM signals.  
The neural network-based modulation classifiers \cite{liu2023towards, sathyanarayanan2023rml22} study how environmental change affects classification performance for only the single-carrier signals, not OFDM signals.

\textbf{Related work on sniffing OFDM signals:} One approach to modulation identification for spectrum sensing uses sniffing of control information which notifies the RX about modulation and coding formats. The work~\cite{kumar2014lte, bui2016owl, falkenberg2019falcon, hoangLtesniffer} attempts to overhear Long Term Evolution (LTE) signals. LTEye~\cite{kumar2014lte} and OWL~\cite{bui2016owl} decode PHY DL control channel (PDCCH) data for LTE network monitoring. LTESniffer~\cite{hoangLtesniffer} decodes sniffed both user and control data using the PDCCH decoder FALCON~\cite{falkenberg2019falcon}. FALCON overcomes the limitation of LTEye and OWL, which require more than 97\% decoding accuracy. 
In LTE, the starting symbol of the PDCCH is always the first symbol in a slot. This is different from 5G, where the PDCCH starting symbol can be any symbol in a slot and its information is notified by radio resource control (RRC) signaling. Accordingly, it is not straightforward to modify the LTE PDCCH sniffer for 5G. Eavesdropping PDCCH data of 5G signals~\cite{ludant20235g} applies to 5G signals with diverse configurations. Still, it is vulnerable to configuration changes since it takes a few minutes to learn a new PDCCH configuration. 
The authors of~\cite{li2020case} study sniffing Wi-Fi probe request packets, which is for mobile devices to broadcast the existence of themselves. They build a hardware model for a sniffer and test with real Wi-Fi probe request packets. However, the probe request packets are simpler in format than those for user data communication. Thus, it is not straightforward to deploy this approach to our setting.

To summarize, the main contributions of the paper are:
\begin{itemize}
\item \textbf{OFDM parameter estimation for up-to-date protocols:} We have applied the OFDM parameter estimation method with CAF~\cite{punchihewa2011blind} to Wi-Fi~6 and 5G signals to estimate SCS and CP length.
\item \textbf{Feature extraction without symbol-level synchronization:} Only with estimated values of SCS and CP length, our system builds the features characterizing modulation of OFDM signals. The proposed feature extraction algorithm is designed to be resilient to symbol-level synchronization errors caused by using CP only.
\item \textbf{Modulation classification without control information:} For spectrum sensing, control information might not be accessible. We show that the proposed classification system robustly works with diverse configurations with the evaluation of hardware-generated data without knowledge of the information.
\end{itemize}

\vspace{-5pt}
\section{System Objective}
\begin{figure}
\centering
 \mbox{\subfloat[]{\label{subfig:sysModel}
       \includegraphics[width=.55\columnwidth]{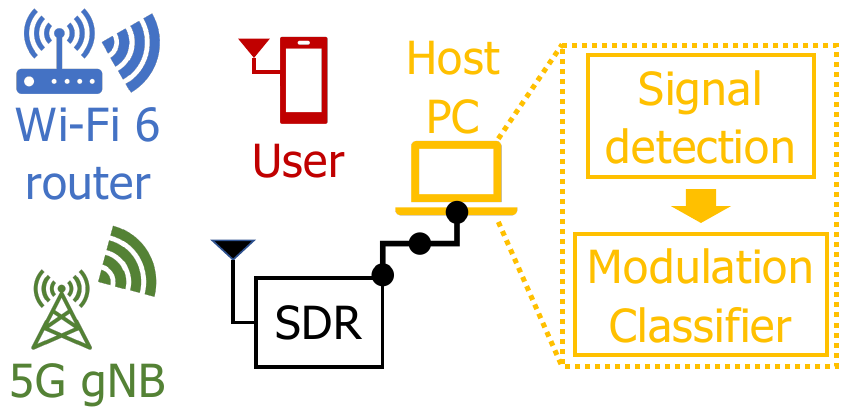}}}
 \mbox{\subfloat[]{\label{subfig:sensing}
       \includegraphics[width=.41\columnwidth]{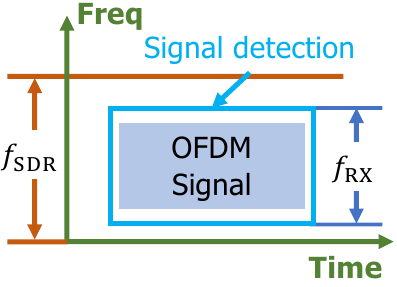}}}
 \caption{(a) To capture DL Wi-Fi~6 and 5G signals and (b) Spectrum sensing scenario using USRP N310.}\label{fig:sys_model}
\vspace{-15pt}
\end{figure}
 
\noindent We aim to build a modulation classifier using IQ samples of SISO Wi-Fi~6 and FR1 5G DL signal for spectrum sensing. The system scenario is described in Fig.~\ref{subfig:sysModel}. There is a Wi-Fi~6 or 5G TX transmitting its signal to an RX. SDR continuously senses the spectrum by generating IQ samples with sampling rate $\fSDR$ and transfers those samples to the host PC. In the host PC, there is a signal detection algorithm and a modulation classifier. Using IQ samples generated from SDR, the signal detection algorithm detects the duration and frequency band where the OFDM signal is located and extracts IQ samples corresponding to the detected OFDM signal, described as the blue rectangle in Fig.~\ref{subfig:sensing}. We assume the accurate signal detection of Wi-Fi~6 or 5G signals and a single modulation scheme is used for data communication in one detected OFDM signal.

The IQ samples from SDR sampled with rate $\fSDR$ are resampled to $\fRX$, 20~MHz. We only consider Wi-Fi~6 signals with 20~MHz channel bandwidth and 5G signals with a PDSCH bandwidth from 15 to 20~MHz. Thus, a 20~MHz sampling rate can let the resampled IQ sequence encompass the OFDM signal in our scenario. Extending the analysis to different transmission bandwidth ranges is straightforward. These resampled IQ samples, denoted by $y[n]$, are taken as inputs of the feature extraction algorithm, as elaborated in Sec.~\ref{sec:algo} in detail.

\begin{table}[]
\centering
\caption{Variable definitions}
\label{tab:var}
\begin{tabular}{c|c}
\toprule
Variable & Definition (unit)
\\ \midrule
$\fTX$ & TX sampling rate (Hz) \\ \hline
$\fRX$ & Sampling rate of a system input sequence (Hz) \\ \hline
$\dFSCS$ & Subcarrier spacing (Hz) \\ \hline
$\TIFFT$ & IFFT sequence duration (s) \\ \hline
$\NFFT$ & FFT size used to generate IFFT sequence \\ \hline
$\TCP$ & CP duration (s) \\ \hline
\multirow{2}{*}{$\NCP$} & Number of time samples \\ & in CP for one OFDM symbol \\ \hline
\multirow{2}{*}{$y[n]$} & Received time-domain sequence after \\ &resampling to 20~MHz \\ \hline
\multirow{2}{*}{$y'[n]$} & 5G time-domain sequence after \\ & resampling to 30.72~MHz \\ \hline
\multirow{2}{*}{$y^s[n]$} & Received time-domain IFFT sequence \\ & for the $s$th OFDM symbol \\ \hline
\multirow{2}{*}{$Y^s[k]$} & Received symbol in subcarrier $k$ \\ & for the $s$th OFDM symbol \\ \hline
$(\mathcal{S}\times\mathcal{S})$ & Number of bins in a 2D histogram \\
\bottomrule
\end{tabular}
\vspace{-10pt}
\end{table}

\vspace{-5pt}
\subsection{Wi-Fi~6 PHY layer}
\begin{table*}[]
\centering
\caption{Parameters for different formats of Wi-Fi}
\label{tab:param_wifi}
\begin{tabular}{c|cccc}
\toprule
 & \textbf{Non-HT format}                                               & \textbf{HT format}                                                   & \textbf{VHT format}                                                            & \textbf{HE format}                                                                      \\ \midrule
\textbf{$\TIFFT$} & $3.2\,\mu\mathrm{s}$ & $3.2\,\mu\mathrm{s}$ & $3.2\,\mu\mathrm{s}$ & $12.8\,\mu\mathrm{s}$ \\ 
\textbf{$\TCP$}                                                      & $0.8\,\mu\mathrm{s}$                                                         & $\{0.4, 0.8\}\,\mu\mathrm{s}$                                                & $\{0.4, 0.8\}\,\mu\mathrm{s}$                                                          & $\{0.8, 1.6, 3.2\}\,\mu\mathrm{s}$s                                                              \\ 
\textbf{Modulations}                                                         & \begin{tabular}[c]{@{}c@{}}BPSK, QPSK,\\ 16QAM, 64QAM\end{tabular} & \begin{tabular}[c]{@{}c@{}}BPSK, QPSK,\\ 16QAM, 64QAM\end{tabular} & \begin{tabular}[c]{@{}c@{}}BPSK, QPSK, 16QAM,\\ 64QAM, 256QAM\end{tabular} & \begin{tabular}[c]{@{}c@{}}BPSK, QPSK, 16QAM,\\64QAM, 256QAM, 1024QAM\end{tabular} \\ \bottomrule
\end{tabular}
\vspace{-12pt}
\end{table*}

\noindent Wi-Fi 6 supports the high-efficiency (HE) transmission format as well as earlier formats, which are non-high throughput (non-HT), high throughput (HT), and very high throughput (VHT) formats. Table~\ref{tab:param_wifi} summarizes the parameters that configure the payload of the Wi-Fi frame for each Wi-Fi format. In HE format, given channel bandwidth, the number of subcarriers is increased because the SCS (denoted as $\dFSCS$) is one-fourth of that of the previous transmission formats. Over time, the Wi-Fi standard has evolved and several options for the CP duration are available.

\vspace{-5pt}
\subsection{5G DL PHY layer}

\begin{figure}
\vspace{-10pt}
\centering
  \mbox{\subfloat[]{\label{subfig:5G_PB}
       \includegraphics[width=.27\columnwidth]{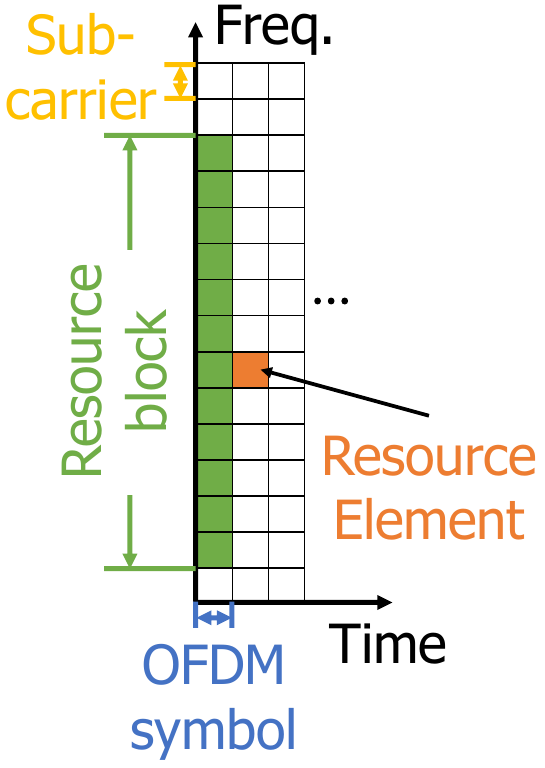}}}
 \mbox{\subfloat[]{\label{subfig:5G_frame}
       \includegraphics[width=.63\columnwidth]{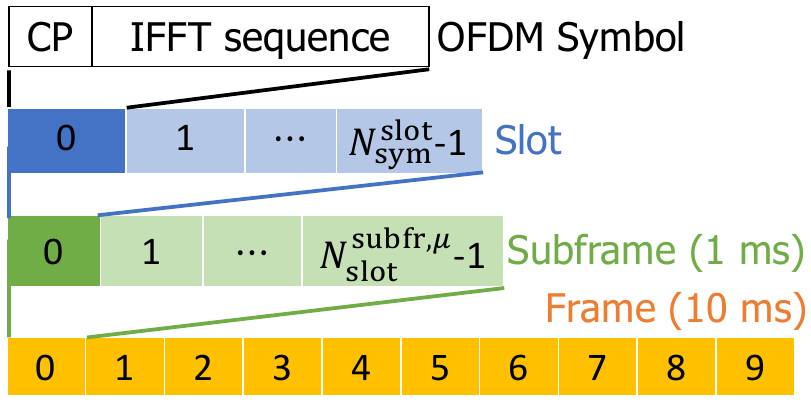}}}
\caption{5G Resource structure: (a) Resource grid and (b) Frame structure.}\label{fig:5G_str}
\vspace{-12pt}
\end{figure}

\begin{table*}[]
\centering
\caption{5G frame structure parameters}
\label{tab:param_5G}
\begin{tabular}{c|cccc}
\toprule
\textbf{\{SCS (kHz), CP option\}}       & \textbf{\{60, Normal\}} & \textbf{\{60, Extended\}} & \textbf{\{30, Normal\}} & \textbf{\{15, Normal\}} \\ \midrule
\textbf{$\TIFFT$}      & $16.17\,\mu\mathrm{s}$            & $16.67\,\mu\mathrm{s}$            & $33.33\,\mu\mathrm{s}$                   & $66.67\,\mu\mathrm{s}$        \\ 
\textbf{\{Short, long\} $\TCP$} & $\{1.17, 1.69\}\,\mu\mathrm{s}$ & $\{4.17, -\}\,\mu\mathrm{s}$ & $\{2.34, 2.86\}\,\mu\mathrm{s}$ & $\{4.69, 5.21\}\,\mu\mathrm{s}$                    \\ 
\textbf{$\NFFT$ when $\fTX=$~30.72~MHz} & 512 & 512 & 1024 & 2048 \\ 
\textbf{\{Short, long\} $\NCP$ when $\fTX=$~30.72~MHz} & \{36, 52\} & 128 & \{72, 88\} & \{144, 160\}\\
\bottomrule
\end{tabular}
\vspace{-12pt}
\end{table*}

\noindent The 5G downlink (DL) resource structure and its associated terminology is illustrated in Fig.~\ref{fig:5G_str}. A resource element (RE), illustrated in Fig.~\ref{subfig:5G_PB}, represents the smallest unit that carries data, encompassing a single OFDM symbol in the time domain and a single subcarrier in the frequency domain. A resource block (RB) is the smallest radio resource that can be allocated and refers to one OFDM symbol in the time domain and 12 subcarriers in the frequency domain. 

Fig.~\ref{subfig:5G_frame} shows the 5G frame structure in the time domain. An OFDM symbol in 5G is comprised of both a CP and an IFFT sequence. The number of symbols within a single slot ($N_{\rm{sym}}^{\rm{slot}}$) varies following the CP length. There are normal and extended CP options in the transmission format. When a normal CP is used then $N_{\rm{sym}}^{\rm{slot}}=14$, otherwise $N_{\rm{sym}}^{\rm{slot}}=12$. The SCS, the distance between two adjacent subcarriers in OFDM systems, determines the number of slots within a single subframe, $N_{\rm{slot}}^{\rm{subfr},\mu}$. $\mu$ represents an SCS option and corresponds to $\dFSCS=15\times2^{\mu}$~kHz. There are five SCS options in 5G, but we consider only three cases, namely 15, 30, and 60 kHz, which are available in FR1. The number of slots in a subframe for each SCS is computed as $N_{\rm{slot}}^{\rm{subfr},\mu}=2^{\mu}$. Finally, one frame of duration 10~ms consists of ten subframes.

The structural parameters that define the 5G frame are listed in Table~\ref{tab:param_5G}. The length of an IFFT sequence, $\TIFFT$, is:
\vspace{-5pt}
\begin{equation}
\label{eq:TIFFT}
	\TIFFT = \NFFT/\fTX = 1/\dFSCS.
 \vspace{-5pt}
\end{equation}
There is a one-to-one correspondence between $\TIFFT$ and $\dFSCS$~\eqref{eq:TIFFT}. Under the normal CP option, CP is longer than that in other symbols, every 0.5 ms, or equivalently, $7\cdot2^{\mu}$ OFDM symbols in OFDM symbol unit, called long CP. There is no long CP in the extended CP option, so $\TCP$ is uniform. The transmission rate of 5G signals is a power of 2 times $15$~kHz and 30.72~MHz is an example of a 5G transmission rate. $\NFFT$ and $\NCP$ values are arranged when $\fTX$ is 30.72~MHz, the value used in our evaluation.


\begin{table*}[]
\centering
\caption{Modulations used for 5G physical channels}
\label{tab:modul_NRPHY}
\begin{tabular}{c|ccccccc}	
\toprule
\textbf{Physical channel} & \textbf{PDSCH}& \textbf{PSS/SSS} & \textbf{PDCCH} & \textbf{CSI-RS} & \textbf{PBCH} & \textbf{PDSCH-PTRS} & \textbf{PDSCH-PTRS} \\ \midrule
\textbf{Modulation}  & \begin{tabular}[c]{@{}c@{}}QPSK, 16QAM, 64QAM, \\ 256QAM, 1024QAM\end{tabular}     & BPSK             & QPSK           & QPSK            & QPSK          & QPSK                                                          & QPSK                                                          \\ \bottomrule
\end{tabular}
\vspace{-15pt}
\end{table*}

In addition to PDSCH, there exist other physical (PHY) channels that serve specific functions although not carrying user data. For instance, PDCCH conveys downlink control information (DCI), which contains information required to decode PDSCH data such as modulation and coding scheme (MCS). Each of these channels utilizes predefined single-type modulation, see Table~\ref{tab:modul_NRPHY}.

Compared to Wi-Fi, which has a predefined configuration of data, pilot, and null subcarriers, 5G resource configuration for PHY channels is flexible. Instead, the 5G system has a network dedicated to exchanging information on how data packets are forwarded, called the control plane, in addition to the network for data transmission, called the user plane. An example of data transferred over the control plane is RRC signals. Information on the starting OFDM symbol of PDCCH and channel state information-reference signal (CSI-RS) is notified to an RX with RRC signals via control plane~\cite{3GPP_38331}.

\vspace{-3pt}
\section{Proposed Algorithm}
\label{sec:algo}
\begin{figure}[t]
\centering{\includegraphics[width=.9\columnwidth]{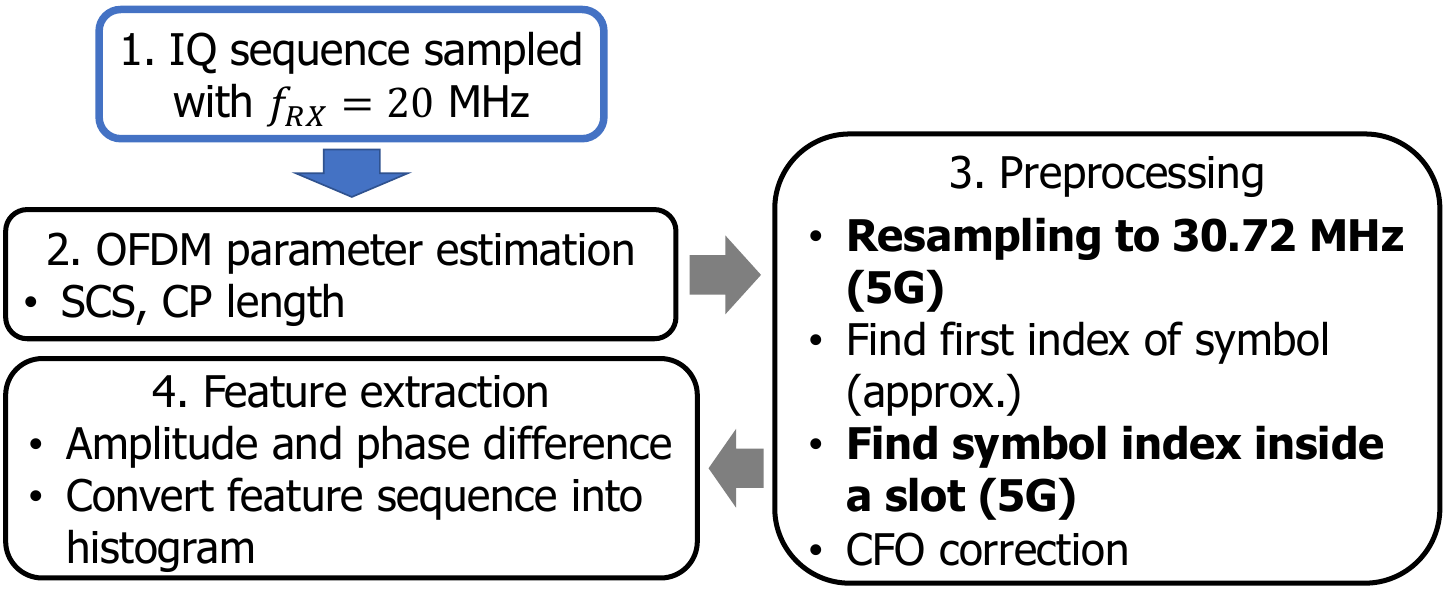}}
\caption{Flow chart of proposed modulation classification algorithm.}\label{fig:flowChart}
\vspace{-10pt}
\end{figure}

High-level procedures to build features characterizing the modulations of Wi-Fi~6 and 5G signals are illustrated in Fig.~\ref{fig:flowChart} and explained in Sec.~\ref{subsec:OFDMParam} and~\ref{subsec:featExt} with additional processing for 5G signals in Sec. \ref{subsubsec:5g}. The 2D histogram is then taken as an input to the neural network model, described in Sec.~\ref{subsec:nn}.

\subsection{OFDM parameter estimation}
\label{subsec:OFDMParam}
Before building the features that characterize modulation, it is necessary to estimate two essential OFDM parameters of OFDM signals, SCS and CP length. To estimate these parameters, we use CAF, a Fourier-series coefficient of the autocorrelation function,
\begin{equation}
\label{eq:CAF}
    \mathcal{R}_{yy}(\alpha, \tau)=\sum_{n=-\infty}^{\infty}\mathcal{R}_{yy}(n,\tau)\Exp{-j2\pi\alpha n}.
\end{equation}
CAF is used to extract a repeated pattern presented in wireless signals~\cite{hong2011dof,cumulant1,punchihewa2011blind}. A variant of the CAF estimator presented in~\cite{punchihewa2011blind} is deployed here,
\begin{equation}
\begin{aligned}
\label{eq:CAFest}
	\displaystyle\hat{\mathcal{R}}_{yy}(\alpha, \ell) = \frac{1}{\mathcal{L}-l-\ell+1}&\sum_{n=0}^{\mathcal{L}-l-\ell}\left\{ \sum_{i=0}^{l-1} y[n+i]y^*[n+i+\ell] \right\} \\ & \times \Exp{-j2\pi\alpha n},
\end{aligned}
\end{equation} 
where $\alpha$ is a cycle frequency and $\mathcal{L}$ is the length of $y[n]$. One sample of our estimator is computed as the autocorrelation with delay $\ell$. It differs from the estimator in~\cite{punchihewa2011blind}, which corresponds to $l=1$ in~\eqref{eq:CAFest}. The increase in the length of a sample sequence $y[n+i]$ aims to make peaks more distinct. We set $l=8$ corresponding to the shortest CP length in our scenario.

CP in OFDM symbols causes a sequence to be repeated at both ends of each symbol. The distance between starting indices of the two repeated sequences at both ends of an OFDM symbol is $\TIFFT$ in time units or $\NFFT(\fRX/\fTX)=\fRX/\dFSCS$ in time sample units. This repetition makes the CAF estimator at $\alpha=0$ have a peak at $\ell=\fRX/\dFSCS$. $\TCP$ is also estimated with the CAF estimator, $\hat{\mathcal{R}}_{yy}(\alpha, \fRX/\dFSCS)$. Since $\sum_{i=0}^{l-1} y[n+i]y^*[n+i+\ell]$ in~\eqref{eq:CAFest} has peaks at period of $\fRX(\TCP+1/\dFSCS)$, it is expected of $\hat{\mathcal{R}}_{yy}\left(\alpha, \fRX/\dFSCS\right)$ to have a large amplitude at $\alpha=1/\{\fRX(\TCP+1/\dFSCS)\}$.

In our scenario, there are five candidates $\ell$ values, $\boldsymbol{\ell}_{C}=\{64, 256, 333, 667, 1333\}$, each corresponding to an IFFT sequence length for a given SCS at $\fRX=20$~MHz. IFFT sequence length is estimated as:
\begin{equation}
	\TIFFT = \ell'/\fRX\;\;\text{s.t.}\;\;\ell'=\operatorname*{arg\,max}_{\ell\in\boldsymbol{\ell}_{C}}\left|\hat{\mathcal{R}}_{yy}(0, \ell)\right|
\end{equation}
When the estimated $\TIFFT$ corresponds to that of Wi-Fi~6 or 60~kHz SCS NR, where multiple CP options are available, CP length is further estimated as:
\begin{equation}
\TCP = \frac{1}{\fRX}\left(\frac{1}{\alpha'}-\ell'\right)\;\;\text{s.t.}\;\;\alpha'= \operatorname*{arg\,max}_{\alpha\in\boldsymbol{\alpha}_C^{\ell'}}\left|\hat{\mathcal{R}}_{yy}(\alpha, \ell')\right|
\end{equation}
where $\boldsymbol{\alpha}_C^{\ell'}$ denotes a set of possible values of $\alpha=1/\{\ell'+(\fRX\cdot\TCP)\}$, given $\ell'$.

\vspace{-3pt}
\subsection{Feature extraction}
\label{subsec:featExt}
The motivation behind our proposed feature extraction lies in the observation that when a sampled time-domain sequence is contained within a single OFDM symbol $s$, the FFT of that sequence yields the original symbols with a phase drift that scales linearly with subcarrier index $k$ and synchronization error $\Delta n$, as shown in
\begin{equation}\label{eq:motiv}
    \begin{aligned}
     Y_{\Delta n}^s[k] & \triangleq \mathcal{F} \left (y^s[n-\Delta n] \right)\\ & = \sum_{n=0}^{\NFFT-1} y^s[n-\Delta n] \Exp{-j2\pi nk/\NFFT}\\
    & = Y^s[k] \Exp{-j2\pi \Delta n k/\NFFT}.
    \end{aligned}
\end{equation}
To build a feature characterizing modulation based on this property, two objectives must be achieved: first, sampling a sequence fully contained in an OFDM symbol, and second, removing the phase drift caused by synchronization errors.

Utilizing the knowledge of $\NCP$ and $\NFFT$, the CP position is determined through autocorrelation analysis,
\vspace{-5pt}
\begin{equation}
\label{eq:CPpos}
    R_{yy}(m, \NFFT) = \frac{1}{\NCP}\sum _{i=0} ^{\NCP-1} {y[m+i] y^*[m+i+ \NFFT]},
\end{equation}
where $m$ is the first index of original sequence of autocorrelation $R_{yy}(m,\NFFT)$. The position of CP is indicated by the peaks in $|R_{yy}(m, \NFFT)|$ since it is expected that $|R_{yy}(m, \NFFT)|$ peaks when $m$ is the first index of CP. To locate a peak, we search for a sample whose amplitude is larger than both of its neighboring samples while ensuring that the minimum distance between two adjacent peaks is 90\% of the OFDM symbol duration (i.e., $(256+64)\times0.9=288$-time samples for HE format with 3.2~$\mu s$ CP), to avoid selecting undesired local peaks. The indices of peaks are denoted as $\{p'_0, \cdots, p'_{S-1}\}$ for $S$ potential OFDM symbols. Using those peaks, the first index of the OFDM symbol is estimated:
\vspace{-3pt}
\begin{equation}
    p = \text{Median}_i \{\text{mod}(p'_i, \NFFT+\NCP)\},
    \vspace{-3pt}
\end{equation}
where $i\in\{0, \cdots j-1\}$. Noise and varying amplitudes of time samples can introduce small errors in the estimated CP position. To reliably sample the sequences contained in a single OFDM symbol, we deploy the sequence $\{y[p+\NCP/2],y[p+\NCP/2+1],\cdots,~y[p+\NCP/2+N-1]\}$. This sequence is entirely within a single OFDM symbol for estimation error of $p$ below $\NCP/2$.

We demonstrated~\eqref{eq:motiv} that $Y_{\Delta n}^s[k]$ exhibits a phase drift, $\Exp{-j2\pi \Delta nk/N}$, while maintaining amplitude $Y^s[k]$. We compute the phase differences between successive potential symbols $s$ and $s+1$ in subcarrier $k$ to build the feature invariant of this phase drift due to synchronization errors as:
\vspace{-3pt}
\begin{equation}\label{eq:phDiff}
    \begin{aligned}
    & \Delta \angle Y_{\Delta n}^s [k] \triangleq \angle Y_{\Delta n}^{s+1}[k] - \angle Y_{\Delta n}^s[k] \\
    & = \angle \left\{Y^{s+1}[k]\Exp{-j2\pi \Delta n k/N} \right\}- \angle \left\{Y^s[k]\Exp{-j2\pi \Delta n k/N}\right\} \\
    & = \angle Y^{s+1}[k] - \angle Y^{s}[k].
    \end{aligned}
    \vspace{-5pt}
\end{equation}
Despite $\Delta n$ unknown, sequences with constant $\Delta n$ are obtained by adjusting the interval between the starting indices of two sampled sequences to be one OFDM symbol. The feature used to identify the modulation type is 
\vspace{-3pt}
\begin{equation}
Y_f^s[k]\triangleq |Y_{\Delta n}^s[k]| \Exp{j\Delta \angle Y_{\Delta n}^s [k]}.
\vspace{-5pt}
\end{equation}
For Wi-Fi~6, the null subcarrier symbols are eliminated by discarding symbols with the $N_{\rm{null}}$ smallest average amplitudes.

In protocol-compliant reception, the Wi-Fi preamble and 5G PDSCH-phase tracking reference signal (PDSCH-DMRS) are deployed for CFO estimation. However, since not accessible to a spectrum sensor, the CP in each OFDM symbol is used for CFO estimation $\Delta f_c$, i.e.,
\begin{equation}
\label{eq:CFOest}
	\angle \left(y[p+\NFFT+i]\cdot y^*[p+i] \right) = 2\pi\Delta f_c/\dFSCS,
\vspace{-3pt}
\end{equation}

where $y[p+i]$ is in CP. We use $i\in\{\lfloor \NCP/4\rfloor, \cdots, \lceil 3\NCP/4 \rceil \}$ so that the sequence $y[p+i]$ are entirely within CP unless estimation error of $p$ exceeds $\NCP/4$. We determine CFO as the average of $\Delta f_c$~\eqref{eq:CFOest} evaluated over multiple OFDM symbols. If the absolute value of the CFO is larger than $\dFSCS/2$, the CFO cannot be accurately estimated due to aliasing. It is discussed in Sec.~\ref{subsubsec:5g}.

\vspace{-5pt}
\subsection{Additional procedures for 5G signal} \label{subsubsec:5g} 
\noindent
To build a modulation feature for 5G, 5G characteristics distinct from those of Wi-Fi, including a different transmission rate, long CP, and flexible usage of subcarriers, should be considered. First, the transmission rate of 5G signals is not $\fRX=20$\,MHz, but is a power of 2 times 15~kHz. Hence, for the signal classified as 5G, we resample the sequence to $\fG=$~30.72\,MHz$=2048\cdot15$\,kHz, the smallest sampling frequency above 20\,MHz. $\NFFT$ and $\NCP$ with 30.72\,MHz sampling rate for each $\dFSCS$ are arranged in the last two rows in Table~\ref{tab:param_5G}.

In the case of the normal CP option, there is a long CP every $T_{\textrm{LCP}}=$~0.5\,ms, which is slightly longer than that of other OFDM symbols. Long CP breaks the assumption of uniform OFDM symbol duration, which is required by the method to find the first indices of OFDM symbols and estimate CFO. Moreover, in building $Y_f^s[k]$, maintaining the fixed interval does not guarantee the constant $\Delta n$ over multiple OFDM symbols. Therefore, long CP also should be located when finding the first index of the OFDM symbol.

\vspace{-7pt}
\begin{algorithm}
\caption{Finding first index of long CP in 5G}\label{alg:findLongCP}
\KwData{($y'[n]$ of length (3~ms + 3 OFDM symbols)), $\mu$}
 $M = 7\cdot 2^{\mu}$, $\NFFT=512\cdot 2^{2-\mu}$, $\NCP=18\cdot 2^{2-\mu}$ 
 \;
 \For{$i=0:5$}{
  $\mathbf{y'_i} \triangleq \{y'[\fG T_{\textrm{LCP}}\cdot i],~\cdots,~y'[\fG T_{\textrm{LCP}}(i+1)+~2(\NFFT+\NCP)+\NCP-1]\}$\;
  Find peaks $\{p'_{i0},\cdots,p'_{i(m+1)}\}$ with $\mathbf{y'_i}$ using $|R_{\mathbf{y'_i}\mathbf{y'_i}}(m,\NFFT)|$~and peak locating function in Sec.~\ref{subsec:featExt}\;
  $p_{ij}=\text{mod}(p'_{ij}, \NFFT+\NCP)$\;
 }
 $\Delta p_j = (\sum_{k=0}^5\{p_{k(j+1)}-p_{k(j-1)}\})/6$ where $j\in\{1,2,\cdots,M\}$\;
 $\{\Delta p_{r_0}, \cdots, \Delta p_{r_{M-1}}\}=\text{sortDescending}(\{\Delta p_j\})$\;
 $\text{symLongCP} = \operatorname*{arg\,max}_{r_q}\text{Var}(\{p_{0r_q}, \cdots, p_{5r_q}\})$ where $q\in\{0,1\}$\;
 $q_{ij}=\begin{cases}
 p_{ij} & \text{if}~j\le \text{symLongCP}\\
 p_{ij}-16 & \text{otherwise}\;
\end{cases}$
 $q = \text{Median}_j(\sum_{k=0}^5q_{kj}/6)$\;
\KwResult{\textbf{IndexLongCP}$=q\!+\!\text{symLongCP}(\NFFT\!+\!\NCP)$}
\end{algorithm}
\vspace{-10pt}

Algorithm~\ref{alg:findLongCP} explains the detailed steps to estimate the first index of OFDM symbol with long CP. $\mathbf{y'_i}$ in line 3 is a sequence cropped to be as long as (0.5~ms + 2 OFDM symbols + $\TCP$). 

In line 4, we find $M+2$ peaks from $\mathbf{y'_i}$ using autocorrelation $|R_{\mathbf{y'_i}\mathbf{y'_i}}(m,\NFFT)|$, where $M$ denotes the number of OFDM symbols in $T_{\textrm{LCP}}$ given $\mu$ and we also compute the autocorrelation at the two symbols at each end. The $M$ average differences between the remainders of two peaks separated by two OFDM symbols modulo OFDM symbol duration, $\Delta p_{j}$, are computed in line 7. We expect that $\Delta p_j$ is the largest when $p_j$ corresponds to long CP. For a more reliable estimation of a long CP, we add a criterion.

In line 10, we choose the two candidates $k_0$ and $k_1$ that give $\Delta p_{k_i}$ the two largest values. We select $k_q$ where the set $\{p_{0k_q}, \cdots p_{5k_q}\}$ has the larger variance between two candidates of $k_q$. This is because we expect that $\{p_{0j}, \cdots, p_{5j}\}$ has the largest variance if $p_{ij}$ corresponds to long CP since long CP makes $|R_{\mathbf{y'_i}\mathbf{y'_i}}(m,\NFFT)|$ a plateau with some width. Using estimated \textbf{IndexLongCP}, we put an additional 16 samples delay at the OFDM symbol with long CP while extracting the feature $Y_f^s[k]$ to maintain uniform $\Delta n$. The number of 16 samples comes from the difference between long CP and non-long CP with a 30.72~MHz sampling rate.

In contrast to Wi-Fi~6 signals, some subcarriers might not be used for transmission amid transmission. If no transmission is made in $Y^s[k]$ or $Y^{s+1}[k]$, their phases are random, and $\Delta \angle Y_{\Delta n}^s[k]$ cannot be the phase difference between two constellation points. Therefore, we set the threshold for the amplitude, denoted as $\beta$, to check whether the RE is being used for transmission. Only when $|Y^s[k]|$ and $|Y^{s+1}[k]|$ are higher than $\beta$, $Y^s[k]$ is used.

The discrepancy between the center frequency of TX and that of received IQ samples of 5G signals might be much larger than for Wi-Fi. In contrast to Wi-Fi, which covers the entire channel bandwidth unless OFDMA is used, PDSCH in 5G might use only the part of channel bandwidth so the center frequency of PDSCH might be different from that used for transmission. Thus, the discrepancy is solely from hardware imperfection in Wi-Fi. For a Wi-Fi link operating at $f_c=5\,\mathrm{GHz}$ and a frequency tolerance of 1~ppm for commercial-off-the-shelf temperature-compensated crystal oscillators~\cite{GTXO-203T} on both sides of the Wi-Fi link, the worst-case CFO is $\Delta f_c = 2f_c \cdot10^{-6}=10$~kHz. However, in 5G, the CFO can escalate to an MHz scale if we consider the center frequency of transmission bandwidth to be carrier frequency. If the method presented earlier in this section is employed, the difference could result in an inaccurate estimation of CFO due to aliasing. Even in the absence of noise, it is only possible to measure $\Delta f_c$ accurately up to $\dFSCS/2$, since $\Delta f_c + z \dFSCS$ cannot be distinguished from each other, where $z\in \mathbb{Z}$. The algorithm makes the corrected CFO a multiple of $\dFSCS$, not a zero.

However, the CFO correction is still deployed for feature extraction. This is because even though this method cannot find the exact CFO, it can recover the orthogonality among subcarriers. The CFO effect in our feature is represented as:
\vspace{-5pt}
$$
\begin{aligned}
	Y_{\Delta n}^s[k] & = \sum_{n=0}^{\NFFT-1} {y[n-\Delta n]\Exp{-j2\pi n \left(\Delta f_c/\fTX + k/\NFFT\right)}} \\
	& = Y^s\left[k+\NFFT\Delta f_c / \fTX\right] \times \\ & \Exp{-j2\pi \Delta n(k/{\NFFT+\Delta f_c / {\fTX})}} \\
  	Y_{\Delta n}^{s+1}[k] & = Y^{s+1}\left[k+\NFFT\Delta f_c / \fTX\right] \times \\ & \Exp{-j2\pi (\Delta n k/{\NFFT+(\Delta n + (\NFFT+\NCP)\Delta f_c / {\fTX})}} \\
\end{aligned}
\vspace{-7pt}
$$
\begin{equation}
\begin{aligned}
   		\Rightarrow \Delta \angle Y_{\Delta n}^s [k] & = \angle Y^{s+1}\left[k+\Delta f_c / \dFSCS\right] \\ & - \angle Y^s \left[k+\Delta f_c / \dFSCS\right] \\ & - 2\pi \Delta f_c (1/\dFSCS+\TCP).
\end{aligned}
\label{eq:CFOorth}
\vspace{-3pt}
\end{equation}
%
To maintain orthogonality of $\angle Y_{\Delta n}^s [k]$ across $k$, $\Delta f_c / \dFSCS$ should be an integer. We have demonstrated that after the CFO correction using CP, the CFO is expressed as $z\cdot \dFSCS$, which renders $\Delta f_c / \dFSCS$ to be an integer. Consequently, the phase of our feature becomes the sum of a phase difference of originally transmitted symbols and a phase caused by the CFO. Since $\Delta \angle Y_{\Delta n}^s [k]$ in~\eqref{eq:CFOorth} contains $\TCP$ term, the CFO effect on $\Delta \angle Y_{\Delta n}^s [k]$ is different when OFDM symbol $s+1$ is an OFDM symbol with long CP. To make the CFO effect uniform in the feature, $\Delta \angle Y_{\Delta n}^s [k]$ where OFDM symbol $s+1$ is an OFDM symbol with long CP is not used for building the feature.

The features may contain the effect of other PHY channels that use modulations other than those used by PDSCH. It is impossible to perfectly filter out the effect because information about which REs were used for which PHY channels is not accessible for spectrum sensors. However, since the modulations of other PHY channels are either BPSK or QPSK, the constellation diagram of the features is only affected by changes in PDSCH modulation. Thus, the distribution of phase differences is still an intrinsic characteristic of PDSCH modulation.

\vspace{-5pt}
\subsection{Neural network classifier}
\label{subsec:nn}
\begin{figure}[t]
\centering \mbox{\subfloat[]{\label{subfig:flowChartNNWlan}
       \includegraphics[width=.58\columnwidth]{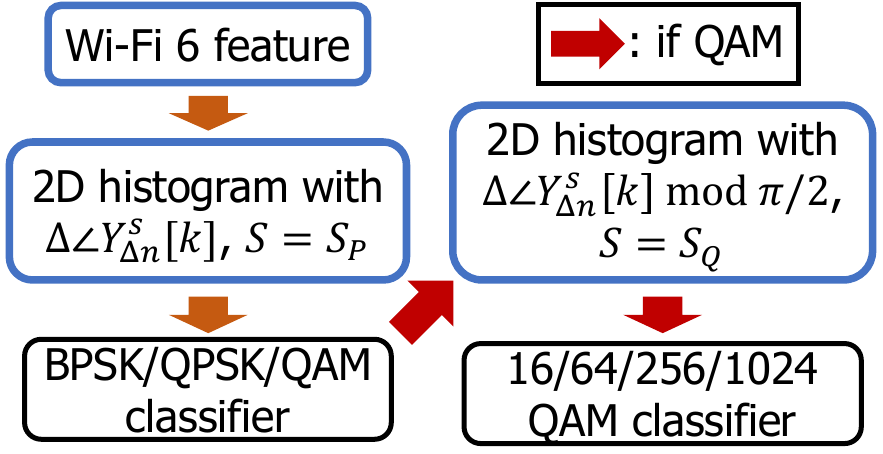}}}
 \mbox{\subfloat[]{\label{subfig:flowChartNN5G}
       \includegraphics[width=.32\columnwidth]{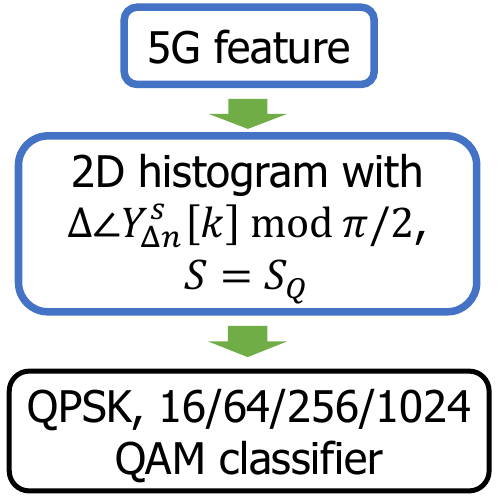}}}
 \caption{Flow chart of proposed classifier system: (a) Wi-Fi~6 and (b) 5G.}\label{fig:flowChartNN}
\vspace{-10pt}
\end{figure}

\begin{figure}[t]
\centering{\includegraphics[width=.95\columnwidth]{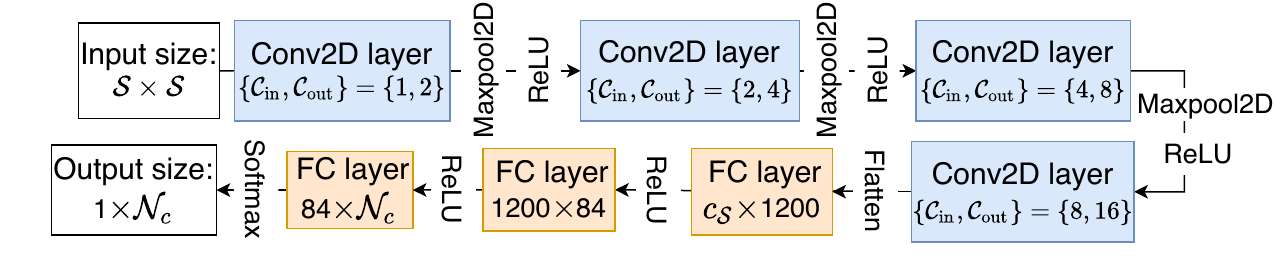}}
 \caption{CNN-based modulation classifier structure. $\mathcal{N}_c$ is the number of modulations a classifier aims to recognize.}\label{fig:CNN}
\vspace{-10pt}
\end{figure}

\begin{table}[]
\vspace{-7pt}
\centering
\caption{DL model parameters}
\label{tab:DL_param}
\begin{tabular}{c|c||c|c}
\toprule
\textbf{Batch size}    & 32               &
\textbf{Learning rate} & $5\cdot10^{-5}$ \\ \hline
\textbf{Epochs}        & 200              &
\textbf{Loss}          & Cross-entropy    \\ \bottomrule
\end{tabular}
\vspace{-15pt}
\end{table}

The obtained feature $Y_f^s[k]$ goes through two preprocessing steps to become input to the classifier: 

1) instead of $\Delta \angle Y_{\Delta n}^s[k]$, $\Delta \angle Y_{\Delta n}^s [k]$ modulo $\pi/2$ is used as a phase of $Y_f^s[k]$. A constellation diagram of every target modulation and corresponding features $Y_f^s[k]$ without noise are rotationally symmetric with $\pi/2$. Thus, $\Delta \angle Y_{\Delta n}^s[k]$ modulo $\pi/2$ is used as a phase of our feature to characterize a modulation. For Wi-Fi~6 signals, BPSK cannot be distinguished from QPSK if $\Delta \angle Y_{\Delta n}^s [k]$ modulo $\pi/2$ is used. Thus, an additional classifier with the original phase as an input is used to distinguish BPSK and QPSK from the high-order QAM modulations, see Fig.~\ref{fig:flowChartNN}. 

2) A 2D histogram of the normalized amplitude of the features $|Y_f^s[k]|/|Y_f^s[k]|_{\textrm{p}99}$, where $|Y_f^s[k]|_{\textrm{p}99}$ denotes 99th percentile of $|Y_f^s[k]|$ in a single data, and the phases $\angle Y_f^s[k]/2\pi$, as an input for the classifier. The histogram value of each bin is computed as:
\vspace{-3pt}
\begin{equation}
\begin{aligned}
Z(u,v) & =\text{The number of}~Y_f^s[k]~\text{s.t.}\\
u/\mathcal{S} & \le |Y_f^s[k]|/|Y_f^s[k]|_{\textrm{p}99} \le (u+1)/\mathcal{S}~\text{and} \\
v/\mathcal{S} & \le \Delta \angle Y_{\Delta n}^s[k] / \phi \le (v+1)/\mathcal{S}.
\end{aligned}
\end{equation}
If $\Delta \angle Y_{\Delta n}^s [k]$ modulo $\pi/2$ is used, $\phi$ is $\pi/2$, otherwise $2\pi$. We normalize histogram value to be classifier input:
\vspace{-3pt}
\begin{equation}
Z'(u,v) =Z(u,v)/\mathcal{Z},
\vspace{-5pt}
\end{equation}
where $\mathcal{Z}$ denotes the number of valid $Y_f^s[k]$ in one data. To remove outliers, $Y_f^s[k]$ whose amplitude is larger than $|Y_f^s[k]|_{\textrm{p}99}$ was not included in the histogram.

The overall structure and the parameter of the classifier with the histogram input are summarized in Fig.~\ref{fig:flowChartNN} and Table~\ref{tab:DL_param}.
The neural network structure used for each classifier is described in Fig.~\ref{fig:CNN}. $\mathcal{C}_\textrm{in}$ and $\mathcal{C}_\textrm{out}$ in Conv2D layers correspond to the number of input and output depth. A $2\times2$ size kernel is used in every Conv2D and Maxpool2D layer. $\mathcal{N}_c$ is the number of modulations that a classifier aims to recognize. For the classifier to identify BPSK and QPSK, the third Maxpool layer is not used, $\mathcal{S}=\mathcal{S}_P$, and $\mathcal{N}_c=3$. The classifier for 5G and for identifying the QAM types for Wi-Fi~6 use $\mathcal{N}_c=5,4$, respectively.

\begin{figure}
\centering
 \mbox{\subfloat[]{\label{subfig:feat_16qam}
       \includegraphics[width=.4\columnwidth]{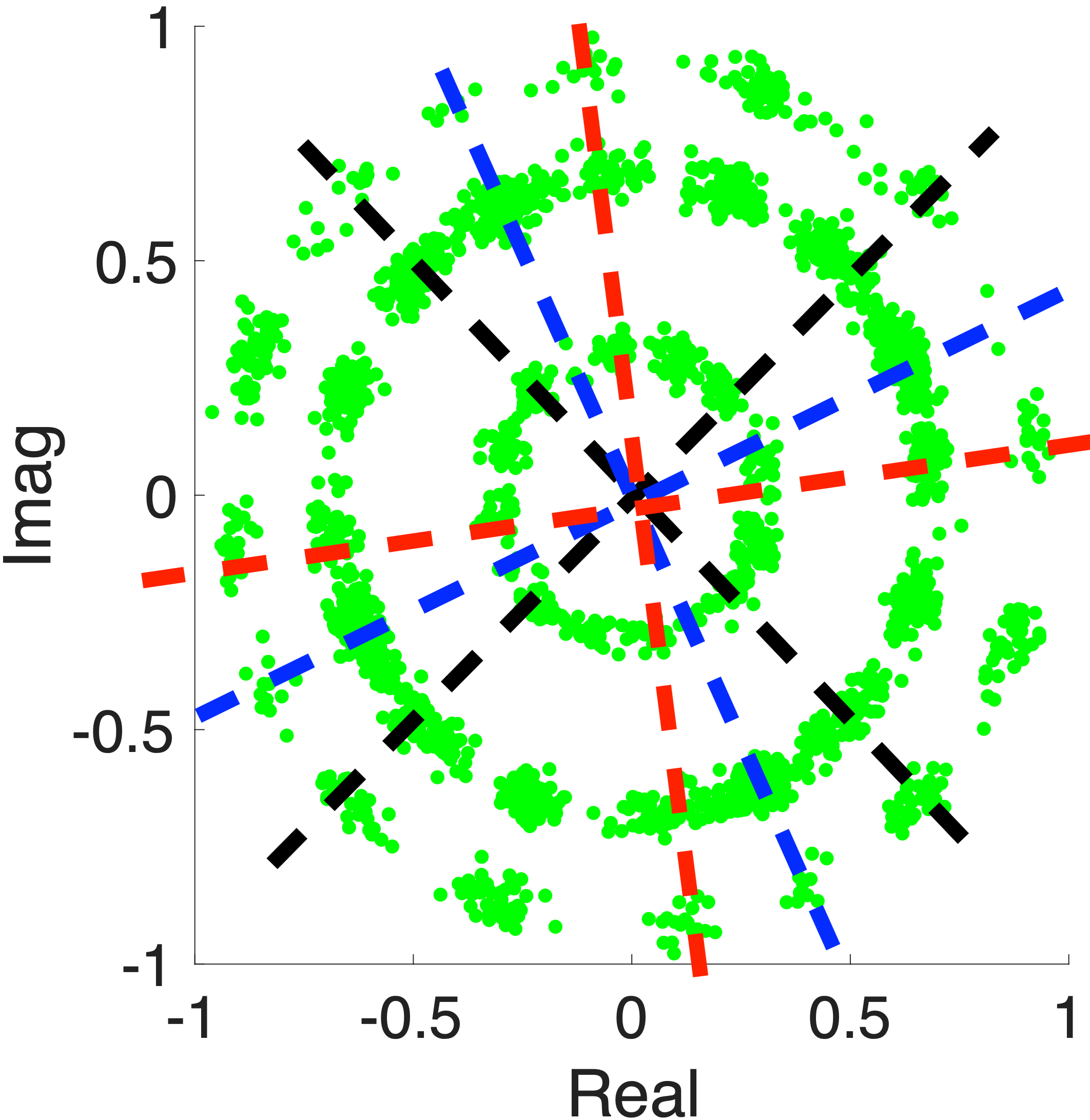}}}
 \mbox{\subfloat[]{\label{subfig:imgFeat_16qam}
       \includegraphics[width=.51\columnwidth]{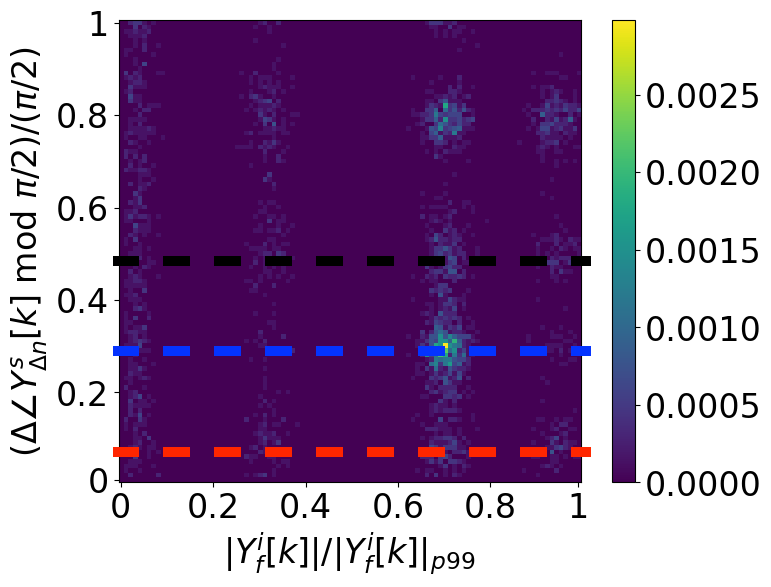}}}
 \caption{Measured 16QAM features at SNR$=25\,$dB with 5G OTA data: 
 (a) Scatterplot of $Y_f^s[k]$ and (b) Corresponding histogram of $|Y_f^s[k]|/|Y_f^s[k]|_{\textrm{p}99}$ and $(\Delta \angle Y_{\Delta n}^s [k]~\textrm{mod}~\pi/2)/(\pi/2)$.}\label{fig:feat}
\vspace{-10pt}
\end{figure}

For 5G 16QAM real-world measured over-the-air (OTA) data, Fig.~\ref{subfig:feat_16qam} shows a scatterplot of the IQ data of $Y_f^s[k]$ and Fig.~\ref{subfig:imgFeat_16qam} the corresponding 2D histogram  with $\Delta \angle Y_{\Delta n}^s[k]$ modulo $\pi/2$. $\angle Y_f^s[k]$ on the red and black dashed lines are the sum of the noise-free phase differences between two 16QAM constellation points and the phase shift caused by CFO. Blue dashed lines are from the phase differences between BPSK or QPSK symbols of the PHY channel other than PDSCH and the shift by CFO. The red, blue, and black dashed lines in Fig.~\ref{subfig:feat_16qam} correspond to the red, blue, and black dashed lines in Fig.~\ref{subfig:imgFeat_16qam}, respectively. Fig.~\ref{subfig:feat_16qam} and Fig.~\ref{subfig:imgFeat_16qam} show that symbols are densely located at the points in the dashed lines, which is consistent with our expectations. 

An advantage of using a histogram is that they are invariant to the length of $Y_f^s[k]$. This enables a neural network with a fixed structure to handle signals of any duration. This property is useful when dealing with 5G features where the number of samples of $Y_f^s[k]$ is unknown due to unused resources. Moreover, in a histogram input, the effect of CFO estimation error caused by aliasing~\eqref{eq:CFOorth} is a movement along the y-axis of the histogram as far as orthogonality of $\angle Y_{\Delta n}^s[k]$ across $k$ holds. The neural network can be trained to identify histogram movements along the y-axis as a single class.

\section{Evaluation}
\label{sec:eval}
\vspace{-5pt}
\subsection{Data collection} 
\begin{figure}
\centering
 \mbox{\subfloat[]{\label{subfig:floormap}
       \includegraphics[width=.22\columnwidth]{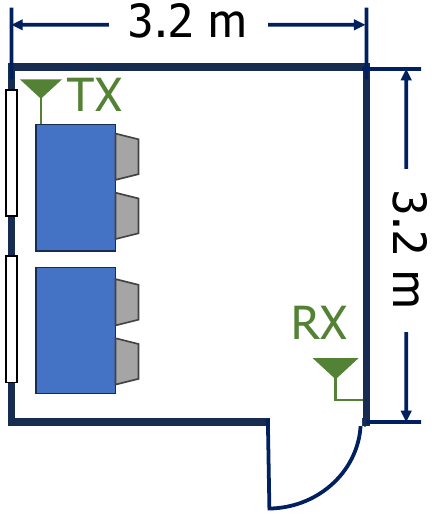}}}
\mbox{\subfloat[]{\label{subfig:ant}
       \includegraphics[width=.30\columnwidth]{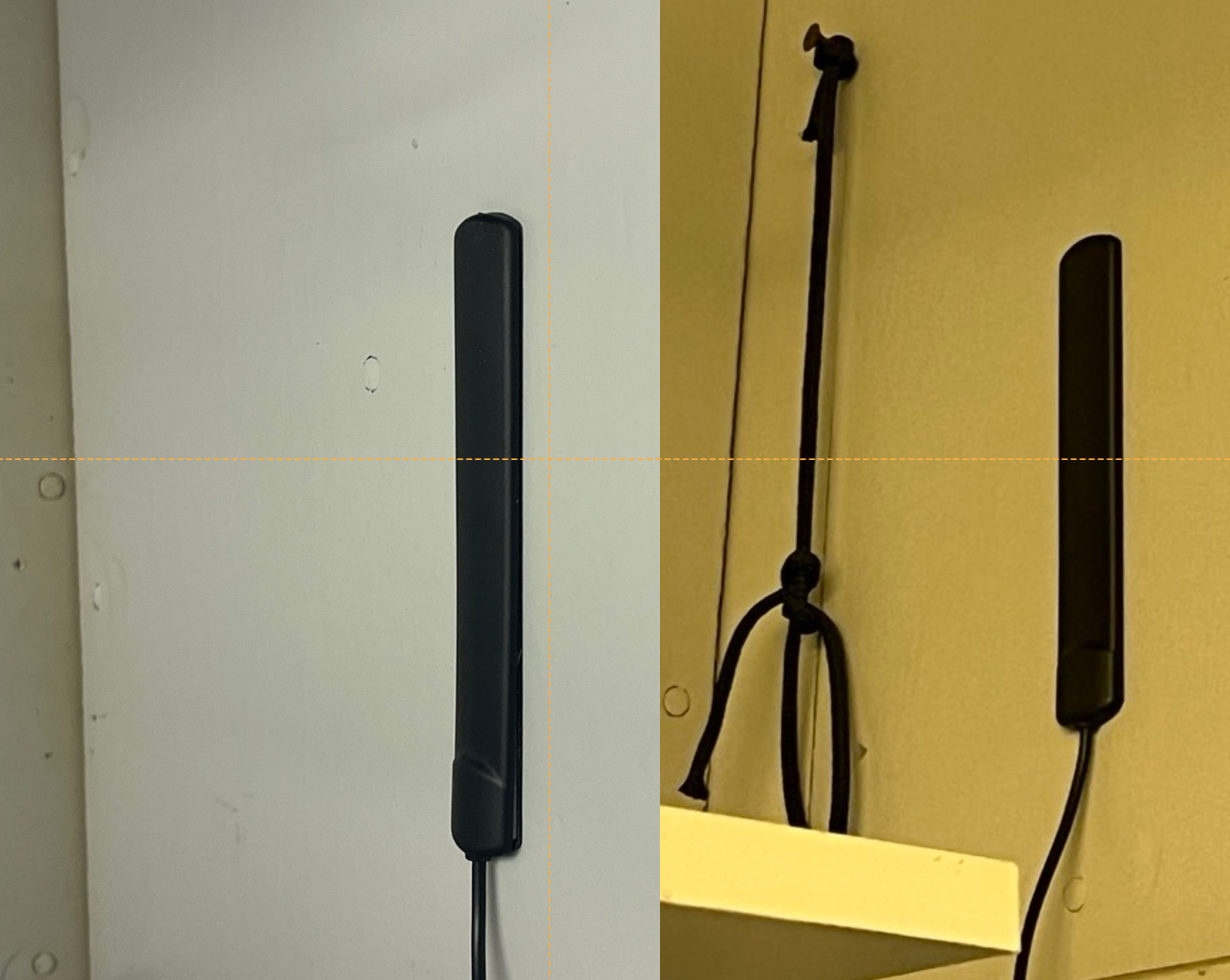}}}
 \mbox{\subfloat[]{\label{subfig:spectro}
       \includegraphics[width=.41\columnwidth]{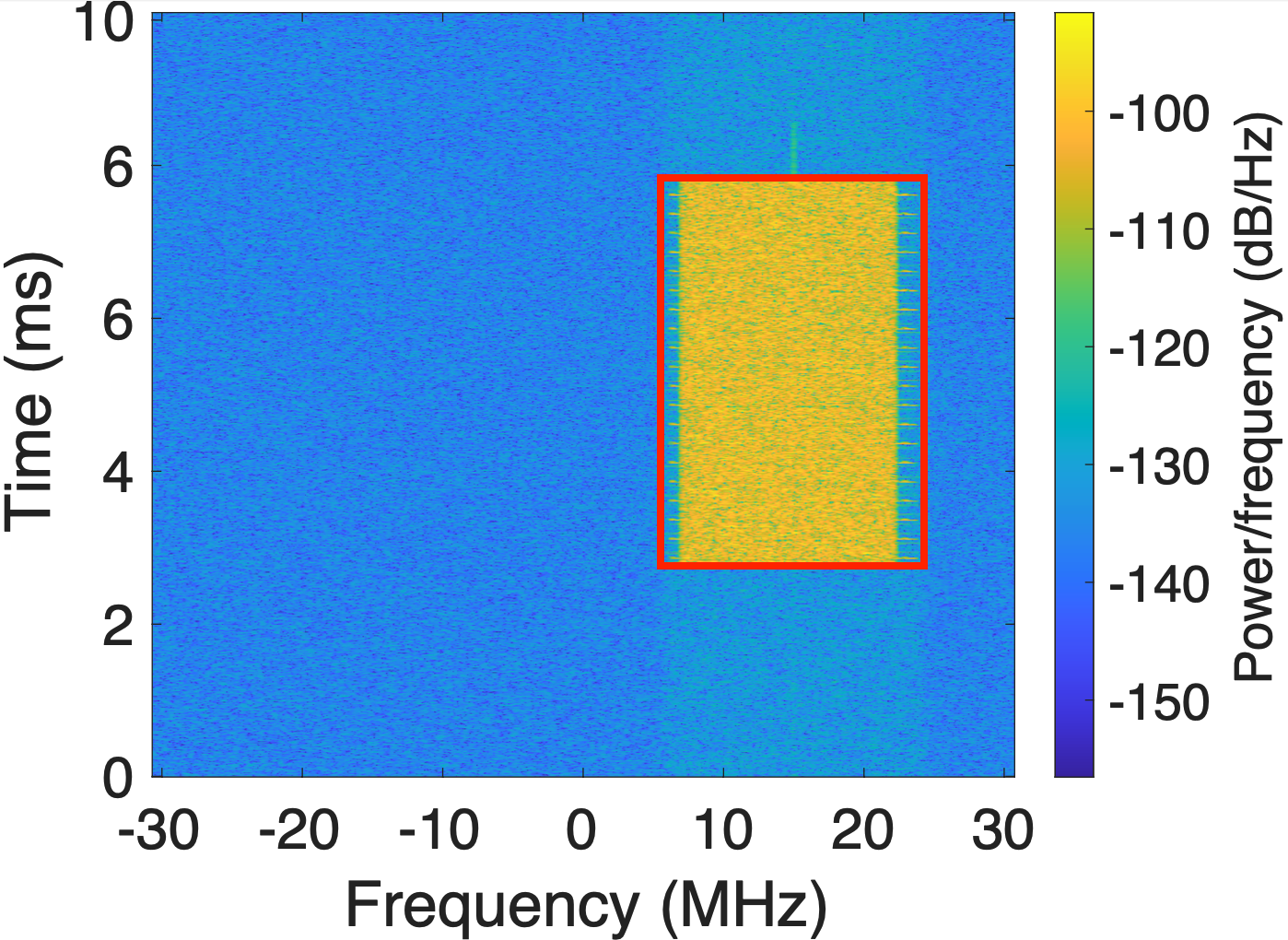}}}
       \vspace{-5pt}
 \caption{OTA data propagation environment: (a) map with TX/RX locations, (b) vertically polarized antennas for TX (left) and RX (right), attached to the wall, and (c) Example spectrogram from data observed by USRP N310.}\label{fig:OTA_env}
\vspace{-10pt}
\end{figure}

\begin{table}[]
\centering
\caption{Data generation parameters}
\label{tab:dataGenParam}
\begin{tabular}{c|c}	
\toprule
\textbf{SNR}     & \begin{tabular}[c]{@{}c@{}}AWGN data: [5, 40]~dB in steps of 5~dB \\ OTA data: [4, 32]~dB in steps of 4~dB\end{tabular}  \\ \hline
\textbf{Carrier frequency} & 2.4~GHz (Wi-Fi~6), 2.6~GHz (5G)\\ \hline 
\begin{tabular}[c]{@{}c@{}}\textbf{The number of} \\ \textbf{\{train, test\} data}\end{tabular}     & \begin{tabular}[c]{@{}c@{}}\{800, 200\} per each\\$(\TIFFT, \TCP, \text{modulation})$ case\end{tabular} \\ \hline
\textbf{$\{\mathcal{S}_P, \mathcal{S}_Q\}$} & \{15, 50\} \\ \hline
\begin{tabular}[c]{@{}c@{}}\textbf{Time duration} \\ \textbf{of each data}\end{tabular} & $400\,\mu\mathrm{s}$ (Wi-Fi~6), 5~ms (5G) \\ \bottomrule
\end{tabular}
\vspace{-10pt}
\end{table}

\begin{figure*}[t]
  \centering
	\subfloat[]{
		\includegraphics[width=0.3\textwidth]{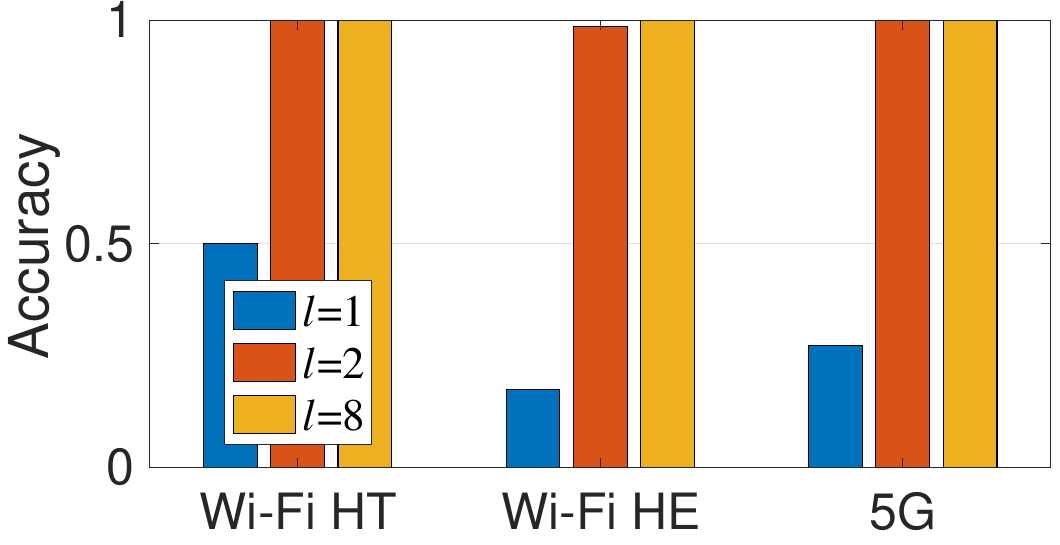}
		\label{subfig:paramEst}
	}
   \centering
	\subfloat[]{
		\includegraphics[width=0.3\textwidth]{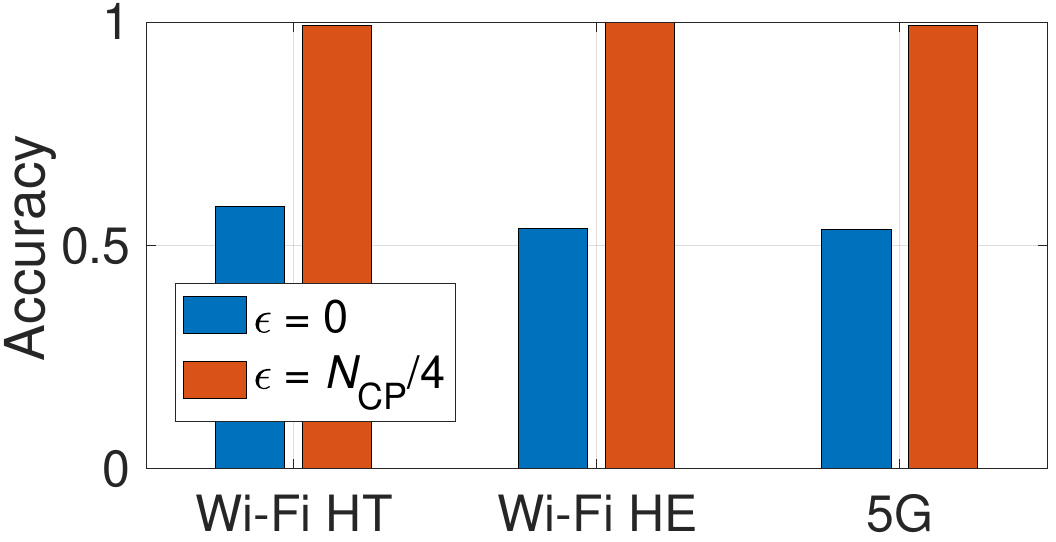}
		\label{subfig:startIndexEst}
	}
    \centering
	\subfloat[]{
		\includegraphics[width=0.3\textwidth]{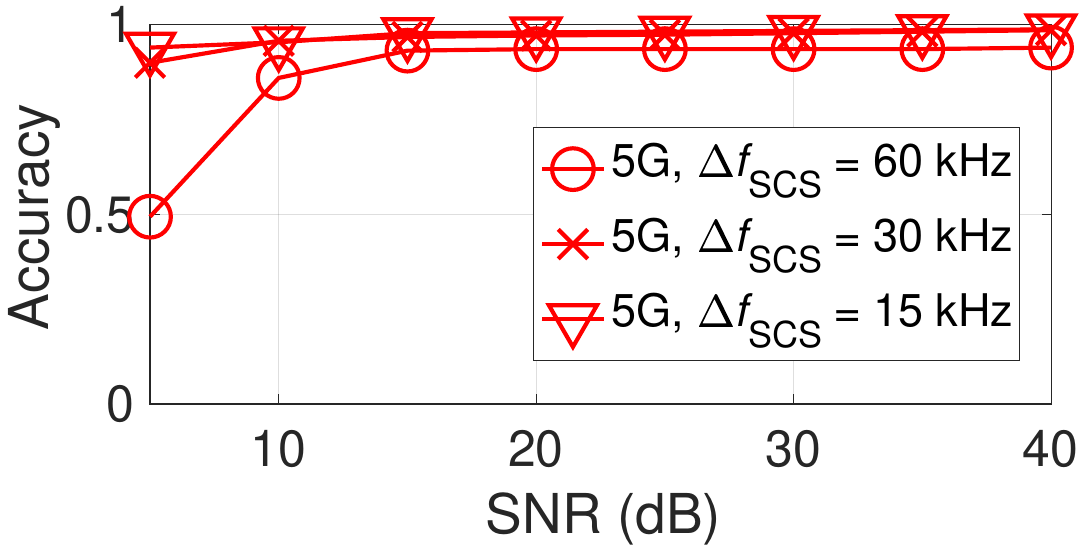}
		\label{subfig:symEst}
	}
 	\caption{Results with synthetic AWGN channel data: (a) Accuracy for estimating $\TIFFT$ and $\TCP$, (b) Accuracy for choosing the first index of CP with acceptable error $\epsilon$, and (c) Accuracy for finding an OFDM symbol with long CP of 5G signals.}
	\label{fig:simPre}
\vspace{-15pt}
\end{figure*}

\vspace{-3pt}
The proposed classifier is evaluated with synthetic data generated from AWGN channel simulations and real-world measured OTA data with the details in Table~\ref{tab:dataGenParam}. MATLAB R2023a WLAN and 5G toolbox~\cite{MATLAB} are deployed to generate the synthetic AWGN dataset. Wi-Fi HT~\cite{ieee802.11-2009} and HE format~\cite{ieee802.11-2021} are used to generate data with $\TIFFT=3.2\,\mu\mathrm{s}$ and $12.8\,\mu\mathrm{s}$ in Wi-Fi~6. For 5G data, every SCS option in FR1, $\mu\in\{0,1,2\}$, is tested. All PHY channels listed in Table~\ref{tab:modul_NRPHY} are included in every 5G data item.

To evaluate whether the performance of the proposed system remains invariant across varying 5G PHY channel configurations, the parameters for allocating REs to PHY channels are set for each data. For example in PDCCH, symbol duration, aggregation level, and starting symbol number are randomly selected. PHY broadcast channel (PBCH), primary synchronization signal (PSS), and secondary synchronization signal (SSS) are included only when $\mu\in\{0,1\}$ since they are not available for $\mu=2$. The other 5G PHY channel parameters are from FR1 test models in \cite{3GPP_38521_4, 3GPP_38141_1}.

Figure~\ref{fig:OTA_env} documents the propagation environment where OTA data are measured. We deploy two networked software-defined radios, USRP N310 \cite{USRP-N310-Reference-Document}, for transmitting and receiving signals OTA. Both TX and RX are in the same room and the distance between TX and RX is $4.52\,\mathrm{m}$, see Fig.~\ref{subfig:floormap}. TX and RX antenna are attached to the wall, see Fig.~\ref{subfig:ant}. Fig.~\ref{subfig:spectro} shows a spectrogram with a 5G signal detected. Utilizing the assumed accurate signal detection, an IQ sequence corresponding to a detected signal (red box in Fig.~\ref{subfig:spectro}) is extracted. After resampling to 20~MHz ($y[n]$), the sequence is taken as an input of the OFDM parameter estimator.

\vspace{-4pt}
\subsection{Building classifier input}

First, to avoid using the Wi-Fi preamble, we remove the first 2000 samples from each data. If the estimated $\TIFFT$ corresponds to those of Wi-Fi~6, an IQ sequence whose length corresponds to 40+2 or 10+2 OFDM symbols is deployed to build $Y_f^s[k]$, starting with a random sample. We need an additional OFDM symbol due to the unknown starting index of an OFDM symbol sequence, $p\in[0,\NFFT+\NCP-1]$. One more symbol is required since phase differences between those of every OFDM symbol and the next one should be computed. $N_{\rm{null}}$ is set to 8 and 32 for Wi-Fi HT and HE, respectively. If the estimated $\TIFFT$ refers to 5G, $y'[n]$ of length (3~ms + 3 OFDM symbols) is used to estimate $p$ and \textbf{IndexLongCP}.

For 5G, the sequence of 14 OFDM symbols is utilized for a classifier input. $\beta$ is set to $|Y_f^s[k]|_\text{p99}/10$ in each input. We also evaluate $Y_f^s[k]$ values as an input to assess how much the histogram input contributes to the performance. In this case, one data input consists of 2240 samples for Wi-Fi~6 or 7900 samples for 5G, which is the average number of feature elements in a single 5G histogram data. We use fixed-duration data for a fair comparison, but the classifier can take the variable length data as input as the obtained feature is processed to a histogram using the algorithms in Sec.~\ref{subsec:nn}. For both input formats, an input with both phases of $\angle Y_{\Delta n}^s[k]$ modulo $\pi/2$ and $\angle Y_{\Delta n}^s[k]$ are evaluated.

\vspace{-4pt}
\subsection{Evaluation results}
\vspace{-4pt}
\subsubsection{AWGN channel data}

\begin{figure*}[t]
  \centering
	\subfloat[]{
		\includegraphics[width=0.29\textwidth]{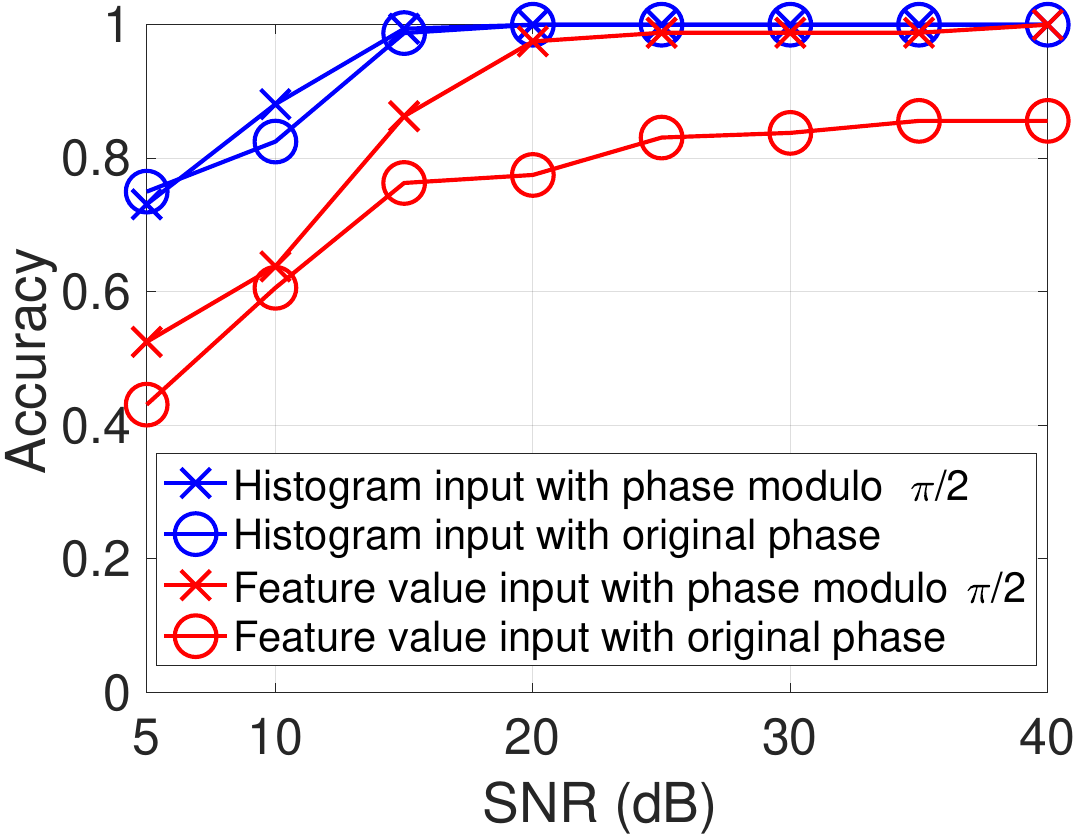}
		\label{subfig:simWifiHT}
	}
   \centering
	\subfloat[]{
		\includegraphics[width=0.29\textwidth]{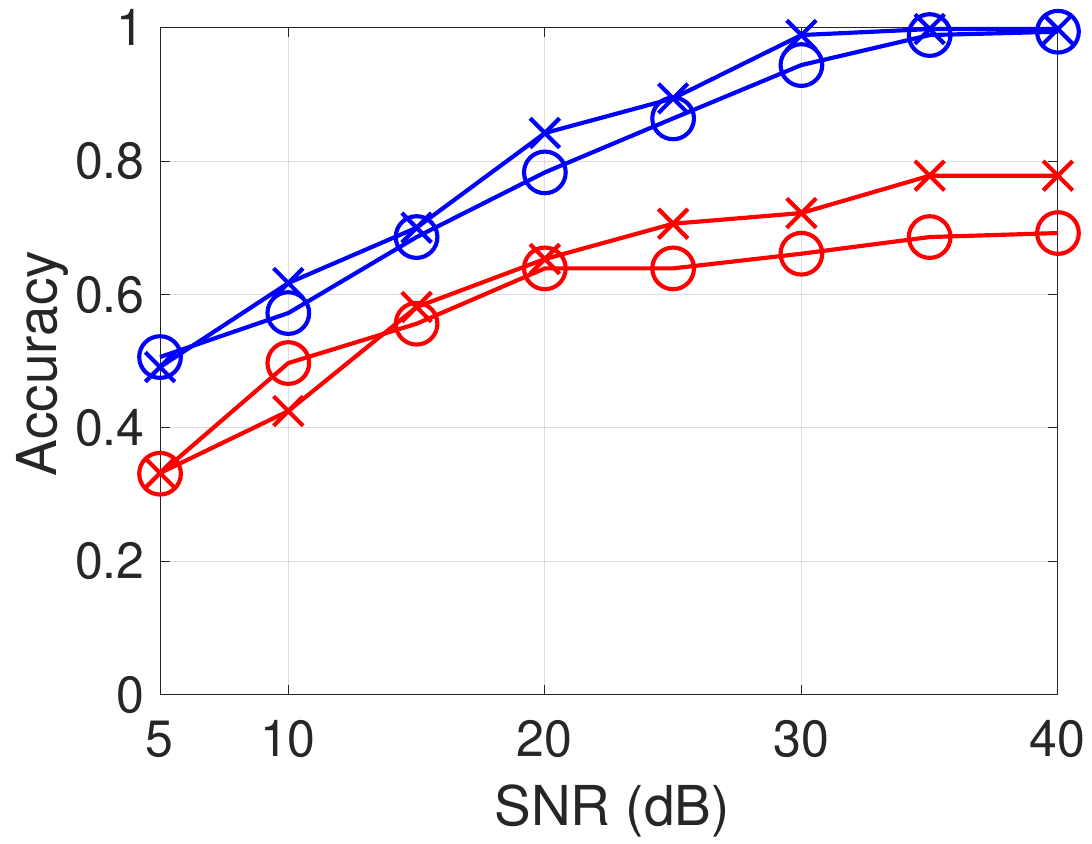}
		\label{subfig:simWifiHE}
	}
    \centering
	\subfloat[]{
		\includegraphics[width=0.29\textwidth]{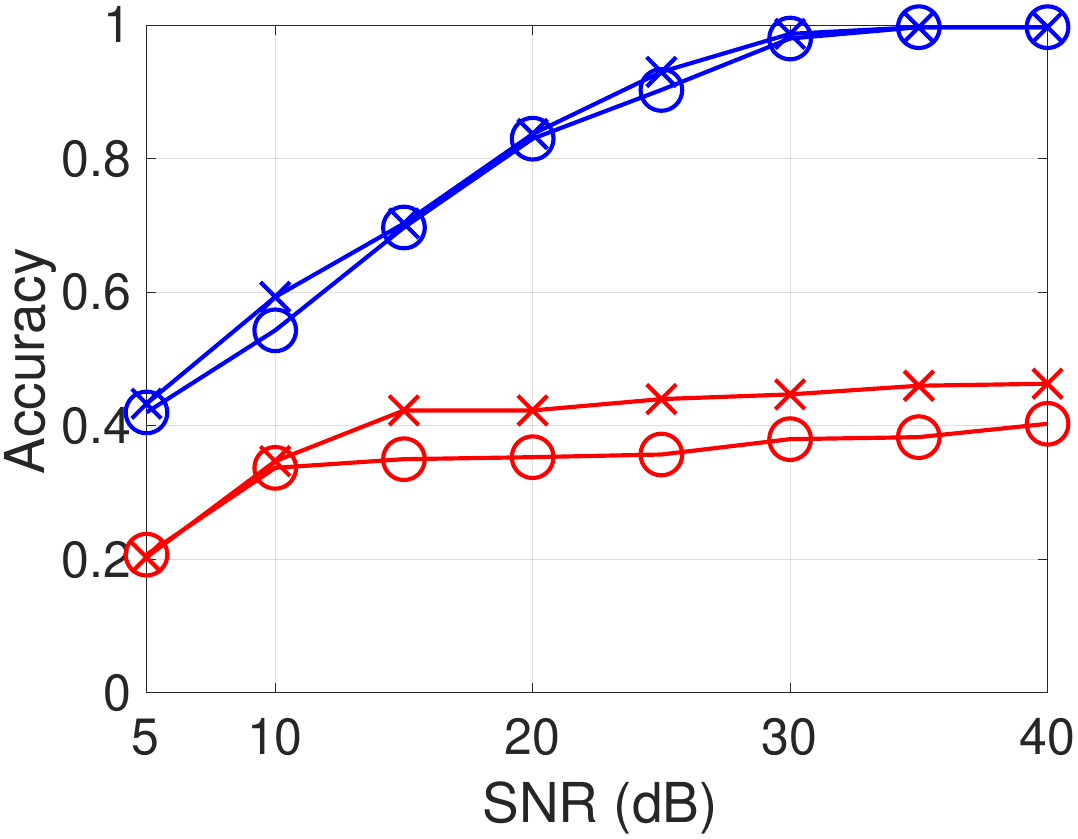}
		\label{subfig:sim5G}
	}
\vspace{-3pt}
 	\caption{Classification accuracy for modulations vs. SNR with synthetic AWGN channel data: (a) Wi-Fi HT, (b) Wi-Fi HE, (c) 5G.}
	\label{fig:simAcc}
\vspace{-15pt}
\end{figure*}

\begin{figure*}[t]
  \centering
   \vspace*{-0.5ex}
	\subfloat[]{
		\includegraphics[width=0.29\textwidth]{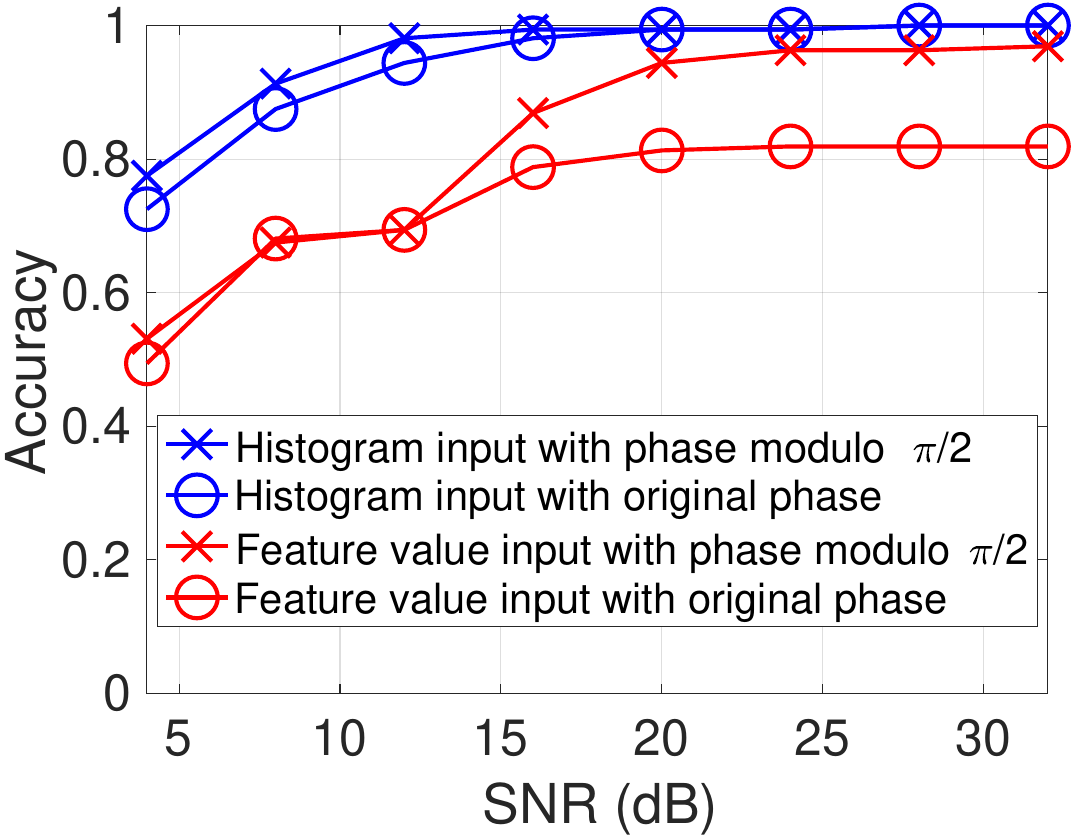}
		\label{subfig:OTAWifiHT}
	}
   \centering
	\subfloat[]{
		\includegraphics[width=0.29\textwidth]{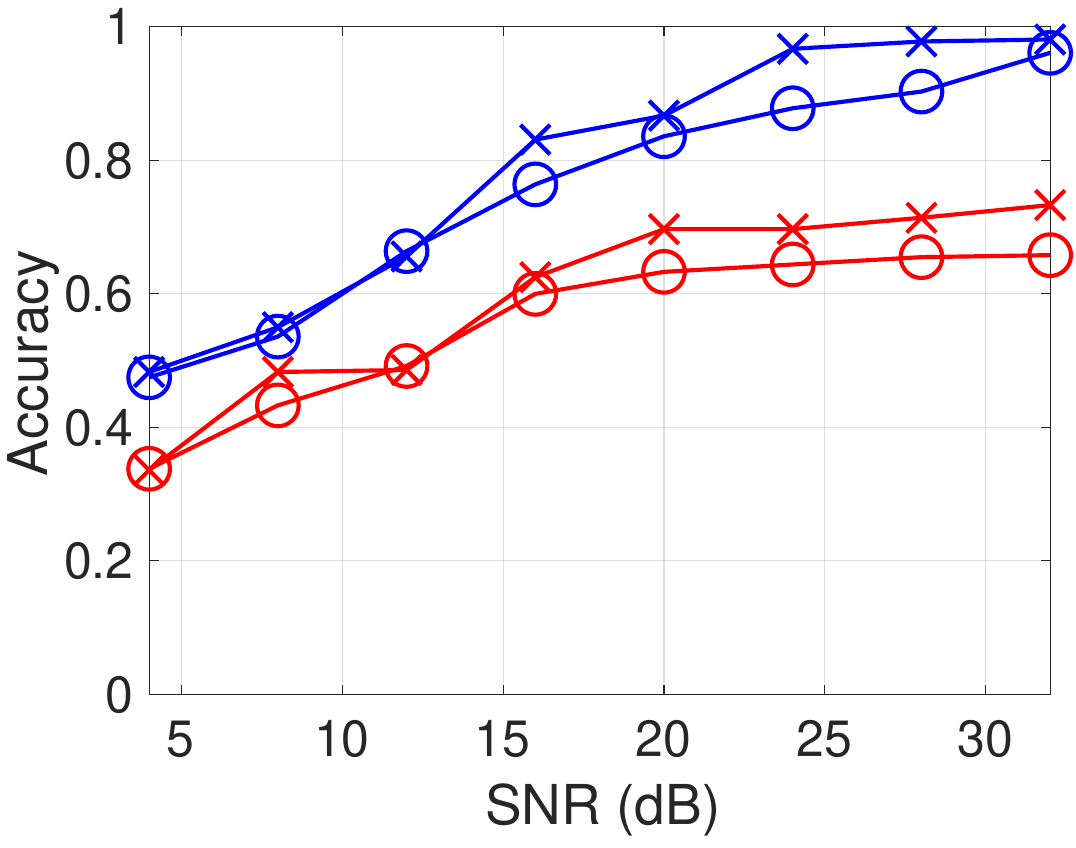}
		\label{subfig:OTAWifiHE}
	}
    \centering
	\subfloat[]{
		\includegraphics[width=0.29\textwidth]{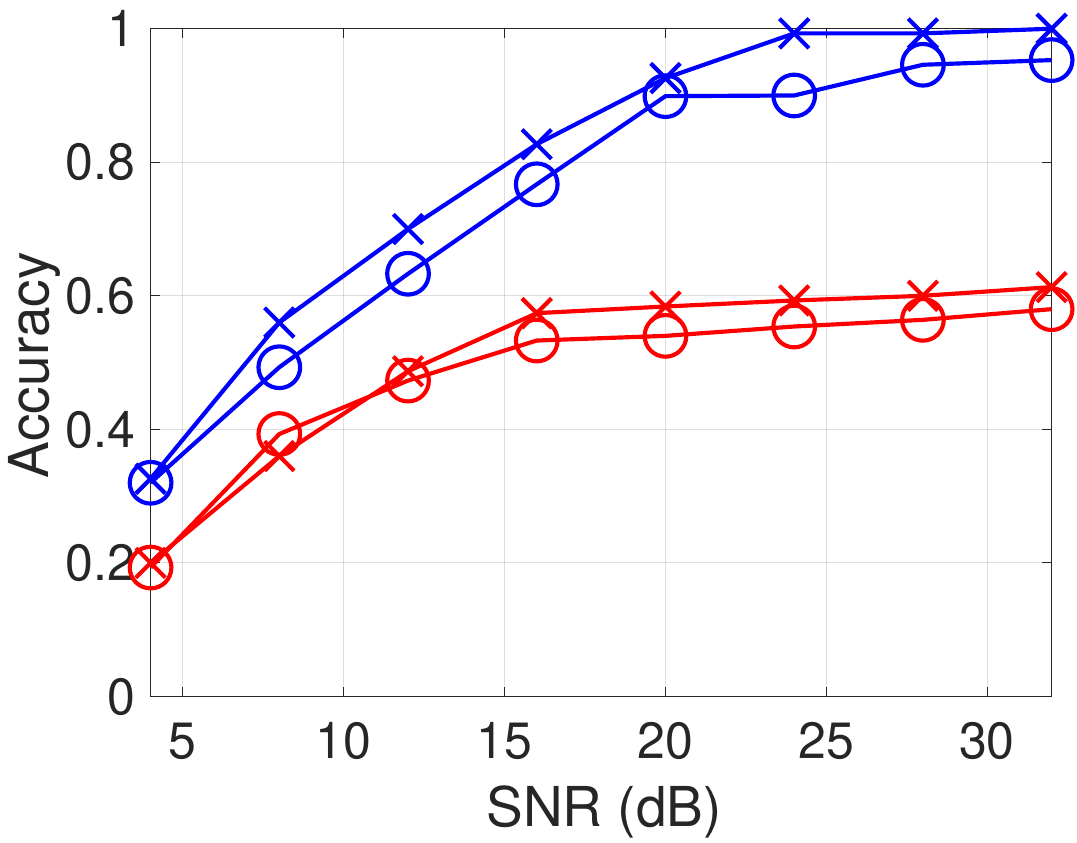}
		\label{subfig:OTA5G}
	}
\vspace{-3pt}
 	\caption{Classification accuracy for modulations vs. SNR with OTA data. (a) Wi-Fi HT, (b) Wi-Fi HE, (c) 5G signals.}
	\label{fig:OTAAcc}
\vspace{-15pt}
\end{figure*}

\begin{figure}[t]
\centering{\includegraphics[width=.9\columnwidth]{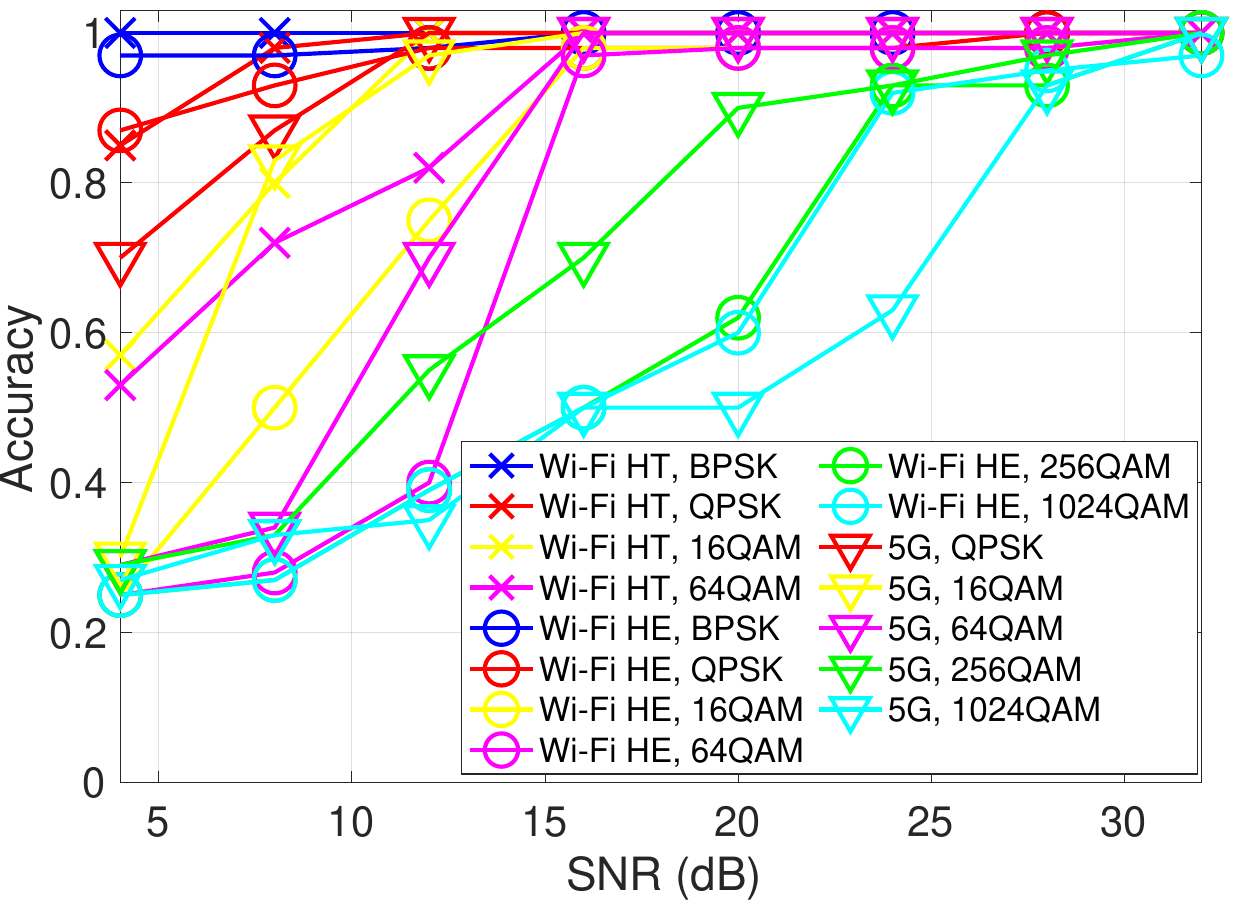}}
\vspace{-10pt}
 \caption{Classification accuracy for modulation with OTA data for each modulation format separately.}\label{fig:accModul}
\vspace{-10pt}
\end{figure}

\begin{figure}[t]
\centering{\includegraphics[width=.9\columnwidth]{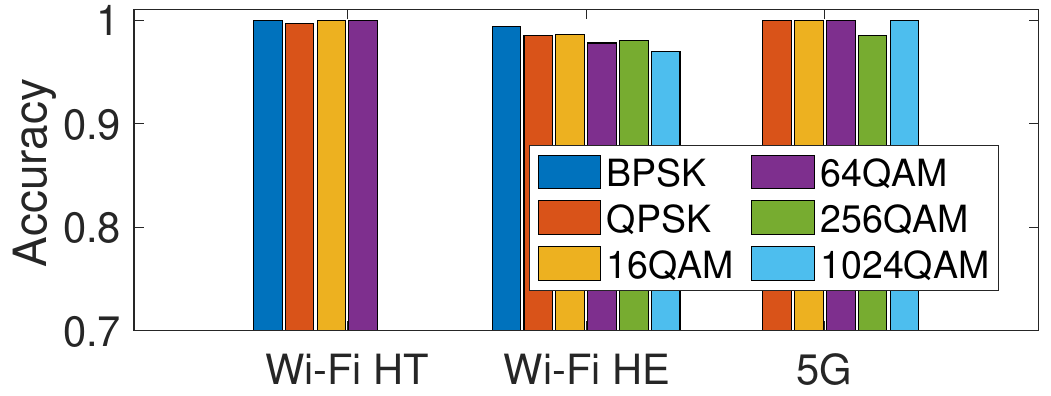}}
\vspace{-8pt}
 \caption{Classifier accuracy with OTA data when SNR exceeds the minimum requirements required for standard-compliant data communication.}\label{fig:accMinSNR}
\vspace{-15pt}
\end{figure}

Results in Fig.~\ref{fig:simPre} are obtained with synthetic AWGN channel data. Fig.~\ref{subfig:paramEst} shows estimation accuracy of the OFDM parameters $\{\TCP, \TIFFT\}$ over different $l$, the length of $y[n+i]$ in CAF estimator~\eqref{eq:CAFest}. Using $l=2,4$ achieves 99\% accuracy for both Wi-Fi~6 formats and 5G and outperforms $l=1$ as used in~\cite{punchihewa2011blind}.
In Fig.~\ref{subfig:startIndexEst}, the estimation accuracy of
\emph{correctly finding} the starting index of an OFDM symbol is shown for the method in Sec.~\ref{subsec:featExt}. \emph{Correctly finding} means that the starting index time is within $\epsilon$ samples tolerance of the true time. In Fig.~\ref{subfig:startIndexEst}, we note that the estimation accuracy for identifying the starting index of an OFDM symbol falls below 60\% for both Wi-Fi~6 formats and 5G. 
When the tolerance is relaxed to $\NCP/4$ time samples, the reported estimation accuracy increases to 99\%.

The accuracy of estimating an OFDM symbol with long CP is shown in Fig.~\ref{subfig:symEst}. Aside from $\dFSCS=60$~kHz, the performance is over 90\% even at low SNR of 5\,dB. Accuracy at $\dFSCS=60$~kHz is low because the period of an OFDM symbol with long CP is larger than the others. The degraded peak detection performance due to the large number of symbols that the peak detection function needs to detect also negatively affects the estimation performance. At $\dFSCS=60$~kHz, 30 peaks should be identified in line 4 of Algorithm~\ref{alg:findLongCP}, which is considerably larger than the 9 or 16 peaks at $\dFSCS=15$~kHz and $\dFSCS=30$~kHz.

Figure~\ref{fig:simAcc} shows modulation classification accuracy with synthetic AWGN channel data. The proposed algorithm with a histogram input with the phases $\Delta \angle Y_{\Delta n}^s [k]$ modulo $\pi/2$ outperforms in all considered cases, except for Wi-Fi~6 at 5\,dB~SNR. The performance gap between using the histogram as classifier input as opposed to using the feature value input increases in Wi-Fi HE and even more so in 5G. This is because the histogram input helps the classifier to discriminate the detailed symbol constellation of high-order modulations.



\begin{table}[]
\centering
\caption{SNR required for data communication with each modulation}
\label{tab:minSNRMod}
\begin{tabular}{c|ccc}
\toprule
\textbf{Modulation}     & BPSK & QPSK & 16QAM  \\ \hline
\textbf{SNR for Wi-Fi~6 (dB)} & 5 & 10 & 16 \\ \hline
\textbf{SNR for 5G (dB)}     & - & 15 & 18 \\ \midrule
\textbf{Modulation}     & 64QAM & 256QAM & 1024QAM \\ \hline
\textbf{SNR for Wi-Fi~6 (dB)} & 22 & 30 & 35 \\ \hline
\textbf{SNR for 5G (dB)}     & 21 & 27 & 30 \\
\bottomrule
\end{tabular}
\vspace{-15pt}
\end{table}

\subsubsection{OTA data} 
The modulation classification accuracy with measured OTA data is in Fig.~\ref{fig:OTAAcc}. The achieved OTA accuracy is similar to the synthetic AWGN channel data: a histogram input with the phases $\Delta \angle Y_{\Delta n}^s [k]$ modulo $\pi/2$ achieves the highest classification accuracy, except for Wi-Fi~6 at 5\,dB~SNR and a larger performance gap for Wi-Fi HE and 5G. 

The classification accuracy of all considered modulation formats with OTA data is in Fig.~\ref{fig:accModul}. For a chosen accuracy, higher modulation orders require higher received SNR. E.g., Wi-Fi HE 16QAM signals have 90\% accuracy if the SNR exceeds 16\,dB, whereas Wi-Fi HE 256QAM requires 24\,dB~SNR. In Fig.~\ref{fig:accMinSNR}, the accuracy of each modulation format is shown
when the SNR satisfies the minimum requirement for standard-compliant data communication. We deploy error vector magnitude (EVM) levels required for data communication with each modulation for Wi-Fi~6 and 5G documentations~\cite{3GPP_38141_1, ieee802.11-2021}. Required SNR values are calculated using the relation between EVM and SNR~\cite{shafik2006extended}. SNR required for the smallest coding rate are chosen for each modulation and chosen values are arranged in Table~\ref{tab:minSNRMod}. For every modulation with both Wi-Fi~6 formats and 5G, accuracy is at least 97\%.

\vspace{-7pt}
\section{Conclusion}
\vspace{-3pt}
Modulation classification of Wi-Fi~6 and 5G signals for spectrum sensing is studied. Simulations show that our classifier which uses SCS and CP length estimates based on the CAF achieves 99\% accuracy. The classifier includes a preprocessing stage that is agnostic to control information, and extracts signal features characterizing modulation schemes insensitive to synchronization errors. 
For 5G signals, the preprocessing also estimates the symbol positions with a long CP. The features are converted to a more suitable form as inputs for the CNN-based classifier. This improves the classification of high-order modulation constellations. The modulation classifier identifies OFDM modulations with 97\% accuracy when the SNR satisfies the requirements for standard-compliant data transmission for each modulation format
with both synthetic AWGN channel data and measured OTA data.
\vspace{-7pt}

\bibliographystyle{IEEEbib}
\bibliography{bibdata,IEEEabrv}
\end{document}